\documentclass[11pt]{article}
\pdfoutput=1
\usepackage{jheppub}

\usepackage{marginnote}

\usepackage{slashed}
\usepackage{color, verbatim}
\usepackage{latexsym}
\usepackage{epsfig}
\usepackage{amsmath}

\usepackage{stackrel}

\usepackage{amssymb}
\usepackage{graphicx}
\usepackage{bm}

\usepackage{hyperref}
\usepackage{epstopdf}
\usepackage{xcolor}
\definecolor{myred}{rgb}{0.7, 0, 0}
\definecolor{myblue}{rgb}{0, 0, 0.7}
\definecolor{mygreen}{rgb}{0.04, 0.7, 0.5}
\hypersetup{colorlinks,citecolor=myred,linkcolor=myblue,urlcolor=myblue,linktocpage=true}

\makeatletter

\newcommand{\be}{\begin{equation}}
\newcommand{\ee}{\end{equation}}
\newcommand{\bea}{\begin{eqnarray}}
\newcommand{\eea}{\end{eqnarray}}

\newcommand{\GW}{\textrm{GW}}
\newcommand{\PBH}{{\rm PBH}}
\newcommand{\DM}{{\rm DM}}

\newcommand{\TeV}{\textrm{TeV}}

\newcommand{\QED}{\textrm{QED}}

\begin{document}

\thispagestyle{empty}

\begin{center}

\begin{center}

  \vspace{0.5cm}

  {\Large\sc
    Complementary Probes of Warped Extra Dimension: \\ \it{Colliders, Gravitational Waves and Primordial Black Holes from Phase Transitions}
  \vspace{0.3cm}
  }  \\

\end{center}

\vspace{1.cm}

\textbf{
Anish Ghoshal$^{\,a}$, Eugenio Meg\'{\i}as$^{\,b}$, Germano Nardini$^{\,c}$, Mariano Quir\'os$^{\,d}$
} \\

\vspace{1cm}

${}^a\!\!$ {\em{Institute of Theoretical Physics, Faculty of Physics,
University of Warsaw, Pasteura 5, PL 02-093, Warsaw, Poland}}

${}^b\!\!$ {\em{Departamento de F\'{\i}sica At\'omica, Molecular y Nuclear and  Instituto Carlos I de F\'{\i}sica Te\'orica y Computacional, Universidad de Granada, Avenida de Fuente Nueva s/n,  18071 Granada, Spain}}

${}^c\!\!$ {\em{Faculty of Science and Technology, University of Stavanger, 4036 Stavanger, Norway}}

${}^d\!\!$ {\em{Institut de F\'{\i}sica d'Altes Energies (IFAE), The Barcelona Institute of  Science and Technology (BIST), Campus UAB, 08193 Bellaterra (Barcelona) Spain}}

\end{center}

\date{\today}

\vspace{2cm}

\centerline{\bf Abstract} 
\vspace{2 mm}

\begin{quote}\small
We study the formation of primordial black holes (PBHs) and stochastic
gravitational waves background (SGWB) produced by the supercooled radion phase
transition (PT) in warped extra-dimension models solving the gauge hierarchy
problem. We first determine how the SGWB and the produced PBH mass and
abundance depend on the warped model's infrared energy scale $\rho$,
and the number of holographic colors $N$. With this finding, we recast on
the plane $\{\rho, N\}$ the current SGWB and PBH constraints, as well
as the expected parameter reaches of GW detectors, as LISA and ET, and the gravitational lensing ones, such as NGRST. On the same plane,
we also map the collider bounds on massive graviton production, and cosmological bounds on the
radion phenomenology. We find that, for $N \sim 10-50$, the considered
PT predicts a PBH population mass in the range $M_{\PBH}\sim(10^{-1} -
10^{-25}) M_{\odot}$ for $\rho \sim (10^{-4} - 10^{8})\textrm{
  TeV}$. 
In the range $\rho \simeq (0.05 - 0.5)$\,GeV, it can explain the recent SGWB hint at nHz frequencies and generate PBH binaries with mass $M_{\PBH}\sim(0.1 - 1 ) M_\odot$ detectable at LISA and ET.
The experimentally allowed mass region where PBHs
can account for the whole dark matter abundance, and are produced with a tuning $\lesssim 10^{-4}$,
corresponds to $10$~TeV
$\lesssim \rho\lesssim$ $10^4$ TeV. 
These PBHs can compensate the lack of natural candidates for dark matter in warped extra dimensional models.
Such a region
represents a great science case where forthcoming and future
colliders like HE-LHC and FCC-hh, gravitational-wave observatories and other PBHs probes play
a key complementary role.
\end{quote}

\vfill




\newpage

\tableofcontents

\newpage

\section{Introduction}
\label{sec:introduction}

The multiple direct observations of gravitational waves (GWs) by the LIGO, VIRGO, and KAGRA (LVK) collaborations have ushered in a new era of astrophysics research and exploration of our Universe. The mergers of neutron stars and stellar-mass black hole binaries have been detected with remarkable precision, and the future looks even brighter. With the improvements in LVK sensitivity and upcoming experiments, the precision of these measurements will further improve, and new classes of GW sources will be detected for the first time. These source classes may encompass: \textit{i)} new transient GW signals, such as those from supermassive black hole mergers, supernova core collapses, or the cosmic string bursts; \textit{ii)}  long-duration (or steady-state) GW sources, such as spinning neutron stars and inspiraling white-dwarf binaries; and \textit{iii)} stochastic GW background (SGWB) arising from the superposition of unresolved astrophysical sources or intrinsically non-localized phenomena predicted in cosmology models.

The detection of a primordial SGWB would provide a unique probe of the early Universe. Primordial GWs travel from the early Universe to the present day with negligible (Planck scale suppressed) interaction, unlike other cosmic relics such as photons and neutrinos. This long free path ensures that the source information carried by primordial GWs remains uncontaminated. So far, the LVK collaborations have searched for SGWBs and found none, setting an upper limit on this class of signals at kHz frequencies~\cite{LIGOScientific:2016jlg, LIGOScientific:2019vic, KAGRA:2021kbb}. At nHz frequencies, instead, the Pulsar Timing Arrays (PTAs) collaborations~\cite{NANOGrav:2023hvm, EPTA:2023fyk, Reardon:2023gzh, Xu:2023wog}  have reported strong evidence of a SGWB, which they can ascribe to astrophysical and/or cosmological sources~\cite{NANOGrav:2023hfp, NANOGrav:2023hvm, EPTA:2023xxk}. Forthcoming and future experiments~\cite{Weltman:2018zrl, Garcia-Bellido:2021zgu, MAGIS-100:2021etm, Badurina:2019hst, AEDGE:2019nxb, Sesana:2019vho, LISA:2017pwj, TianQin:2015yph, Ruan:2018tsw, Kawamura:2020pcg, Corbin:2005ny, Punturo:2010zz, Reitze:2019iox, Aggarwal:2020olq, Berlin:2021txa, Herman:2022fau, Bringmann:2023gba,Valero:2024ncz}  will improve current SGWB measurements by increasing sensitivities around nHz and kHz frequencies as well as extending them to the entire nHz-GHz frequency range.

It is well known that spontaneous symmetry breakings occurring in the early universe may undergo strong first-order phase transitions (FOPTs). During FOPTs, bubbles containing the broken phase nucleate, expand, collide, and percolate. The bubble collisions, along with the motion of the surrounding thermal plasma, are violent, energetic processes that generate GWs~\cite{Witten:1984rs,Hogan:1986qda}. The overall signal is a SGWB with a distinct frequency shape, peaking at the frequency $f_{\rm p} \sim T_0 T/M_{\rm Pl}$ (see Ref.~\cite{LISACosmologyWorkingGroup:2022jok} for a review). In particular, the SGWB from a FOPT at the electroweak (EW) scale peaks at mHz frequencies, within the LISA band, while the SGWB from a FOPT occurring at the EeV scale peaks at kHz frequencies,  i.e., in the frequency band the ground-based GW interferometers are sensitive to. Notably, none of the spontaneous symmetry breakings predicted by the Standard Model (SM) occur via a FOPT~\cite{Kajantie:1996mn, Laine:2015kra, Bhattacharya:2014ara}, making the potential detection of a FOPT SGWB a revolutionary target. Its detection would indeed be striking proof of the existence of beyond the SM (BSM) physics.

BSM physics is expected for several reasons. The SM suffers from severe drawbacks, including the dark matter (DM) puzzle, the baryon asymmetry of the universe, and the electroweak hierarchy problem. The Randall-Sundrum (RS) setup~\cite{Randall:1999ee} is a promising option to solve these problems. This theory involves a mechanism that breaks the conformal symmetry by generating a potential for the radion, the degree of freedom associated with the distance between the electroweak (EW) and Planck branes along the warped space dimension. If this mechanism is realized {\it \`a la} Goldberger-Wise~\cite{Goldberger:1999uk}, the conformal symmetry breaking due to the radion happens via a 
FOPT~\cite{Creminelli:2001th,Randall:2006py,Kaplan:2006yi,Nardini:2007me,Hassanain:2007js,Konstandin:2010cd,Bunk:2017fic,Dillon:2017ctw,vonHarling:2017yew,Baratella:2018pxi,Megias:2018sxv,Agashe:2019lhy,Fujikura:2019oyi,Megias:2020vek,Bigazzi:2020phm,Agashe:2020lfz,Megias:2023kiy,Mishra:2023kiu,Mishra:2024ehr,Barbosa:2024pyn,Agrawal:2025wvf,Gherghetta:2025krk}, 
and the induced warped factor naturally explains the EW-Planck hierarchy. Such a radion FOPT is a key ingredient for solving the baryon asymmetry in RS-inspired models~\cite{Nardini:2007me, Konstandin:2011ds, Bruggisser:2022rdm, Bruggisser:2018mus}. Moreover, it may help address the DM puzzle in RS-like models, as we investigate in the present work for the first time. The idea is that the radion FOPT can trigger the formation of primordial black holes (PBHs)~\cite{Zeldovich:1967lct,Hawking:1971ei,Carr:1974nx} with masses and abundances such that PBHs mimic DM while being compatible with current DM and PBH constraints. For instance, the PBH-to-DM density ratio $f_{\PBH} \equiv \rho_{\PBH} / \rho_{\DM} \simeq 1$ is experimentally allowed when the PBHs have a monochromatic mass distribution in the range~\cite{Carr:2021bzv, LISACosmologyWorkingGroup:2023njw}
\begin{equation}
10^{-16}M_{\odot} \lesssim M_{\PBH} \lesssim 10^{-10}M_{\odot} \,.
\label{eq:windowallowed}
\end{equation}

Various mechanisms acting during FOPTs lead to the density inhomogeneities required for PBH formation~\cite{Hawking:1982ga, Crawford:1982yz, Kodama:1982sf, Hsu:1990fg, Moss:1994iq, Khlopov:1998nm, Lewicki:2019gmv, Liu:2021svg, Hashino:2021qoq, Gross:2021qgx, Baker:2021sno, Kawana:2021tde, He:2022amv, Hashino:2022tcs, Kawana:2022olo, Lewicki:2023ioy, Gouttenoire:2023naa, Salvio:2023ynn, Ai:2024cka}. One such mechanism unavoidably arises in slow, extremely supercooled FOPTs~\cite{Sato:1981bf, Kodama:1981gu, Maeda:1981gw, Sato:1981gv, Hawking:1982ga, Kodama:1982sf, Lewicki:2019gmv, Ashoorioon:2020hln, Kawana:2021tde, Liu:2021svg, Jung:2021mku, Hashino:2022tcs, Huang:2022him, Kawana:2022lba, Kawana:2022olo, Kierkla:2022odc, Kierkla:2023von, Hashino:2021qoq, He:2022amv, Gehrman:2023esa, Lewicki:2023ioy, Gouttenoire:2023naa, Gouttenoire:2023pxh, Baldes:2023rqv, Salvio:2023ynn, Salvio:2023blb, Banerjee:2023qya, Goncalves:2024vkj, Conaci:2024tlc, Cai:2024nln, Arteaga:2024vde}. Its rationale is rather intuitive. When the FOPT is extremely supercooled, the temperature of the thermal bath in the unbroken phase can be so low that the radiation energy density is negligible compared to the vacuum energy density. The FOPT from this vacuum-dominated phase to the broken one thus occurs with huge reheating.
Indeed, in the approximation that reheating is instantaneous, the volume occupied by a just-nucleated bubble (full of broken phase) contains the same energy density immediately before and after the bubble nucleation, which is possible only if a large radiation energy density appears after nucleation. In other words, at the time of nucleation, the broken-phase energy density inside the bubble is equal to that of the unbroken phase where the bubble has appeared. However, this energy density equality no longer holds at later times: inside the bubble, the energy density dilutes over time as radiation, while it remains constant in the unbroken phase where the vacuum energy dominates. This implies that, at a given time during the FOPT, an older bubble has less energy density than a younger one, and the larger their nucleation-time gap, the greater their energy density difference. For this reason, the energy density inhomogeneities are more pronounced if the FOPT is slow, i.e., the times at which the FOPT starts and ends are well separated.

At a quantitative level, the aforementioned PBH formation mechanism is not settled. For the PBH mass distribution, we consider the leading-order contribution yielding a monochromatic mass spectrum. Nevertheless, subleading 
corrections to this result are plausible~\cite{Lewicki:2023ioy}. Moreover, general relativistic numerical simulations of PBH formation~\cite{Shibata:1999zs, Musco:2004ak, Harada:2013epa, Musco:2018rwt, Musco:2020jjb, Germani:2018jgr, Musco:2018rwt, Escriva:2019phb, Escriva:2021pmf, Escriva:2022bwe} do not include the bubble reheating processes, but our estimate of $\mathcal{O}(1)$ overdensities leads us to assume that, to some extent, the collapse should occur. Given these uncertainties, we follow the available established estimates~\cite{Liu:2021svg, Hashino:2021qoq, He:2022amv, Hashino:2022tcs, Kawana:2022olo, Gouttenoire:2023naa, Salvio:2023ynn}, with the understanding that they will need to be replaced with more precise ones when available. In any case, future improvements should not revolutionize the qualitative conclusions we achieve in the present study. 


In models of warped extra dimensions, the radion naturally undergoes a supercooled FOPT~\cite{Baratella:2018pxi,Megias:2020vek,Megias:2023kiy}. In this work, we analyze the phenomenological implications of this FOPT, including the aforementioned PBH formation. For concreteness, we focus on two simple warped setups involving either two or three branes, with the SM fields appropriately localized on one of them. We consider either one or the other scenario depending on the radion breaking scale $\rho$. Specifically, for scenarios where $\rho$ is at TeV or above, we assume the setup with the SM fields localized on the IR brane, as in the original RS scenario~\cite{Randall:1999ee}\footnote{In more complicated (realistic) theories, the SM fields, except possibly the Higgs boson and the right-handed top quark, propagate in the bulk, and a realistic theory of flavor can be constructed. For simplicity, and to stress the main points of our mechanism for the production of PBH and GWs, we assume here that all the SM fields are four-dimensional and propagate on a brane.  The flavor problem then remains as in the SM. Moreover, the precision electroweak data are not in tension with to Kaluza-Klein (KK) states, as we shall see later on.}. For scenarios with $\rho \lesssim 1$ TeV, we instead work with the the SM fields localised on an additional, intermediate brane (between the IR and UV branes) with scale $\rho_T \sim 1$ TeV. In both setups, the only degrees of freedom propagating in the bulk are the graviton and the radion, with well-defined couplings to the SM brane fields. Additionally, the warped factors between the branes naturally explain the branes' scale hierarchies. In both setups, the Goldberger-Wise stabilization mechanism is assumed.

Once the extra dimension is integrated out, our warped setups contain towers of heavy states, the KK modes of all particles propagating in the bulk. They also contain the radion, which typically remains the lightest BSM state and plays a relevant role in collider and early-universe phenomenology. On the one hand, the radion, together with the KK particles, could give rise to resonant signals at the LHC or future colliders, whose so-far null searches impose upper bounds on the couplings between the SM particles and the radion and KK fields~\cite{Arganda:2024nyi}. On the other hand, as previously explained, the radion undergoes a FOPT that potentially generates a detectable SGWB and a PBH population solving the DM puzzle. The overall phenomenology is then very rich and, as we will conclude in the present studies, can be tested by fully leveraging the synergy of present and future GW, PBH, and collider experiments.\\

\textit{The paper is organized as follows.} In Sec.~\ref{sec:Model} we introduce the two warped extra-dimension setups that we investigate. In Sec.~\ref{sec:PhaseTransition} we study the radion FOPT and the key temperatures that characterize it. Section~\ref{sec:PBH} provides the relevant formul$\ae$ for PBH formation and recasts the main PBH observables in terms of the parameters of our warped models. Section~\ref{sec:GW} quantifies the SGWB produced by the radion FOPT and its detection prospects at present and future GW experiments. In Sec.~\ref{sec:collider} we investigate the models' collider signatures due to the couplings between the SM particles, the graviton, and the radion, and we study some astrophysical bounds that constrain the radion lifetime. In Sec.~\ref{sec:conclusions} we compare the parameter reach of the PBH, GW, and collider searches obtained in the previous sections, highlight their synergy and complementarity, and comment on our main results. Finally, in App.~\ref{sec:AppA} we derive the formul$\ae$ for the relevant FOPT temperatures obtained in the thick-wall approximation, while in App.~\ref{sec:AppB} we show how to improve them through a semi-analytical approach.

\medskip

\section{Two-model setups}
\label{sec:Model}
\noindent
We investigate two 5D setups naturally leading to supercooled FOPTs and solving, or at least alleviating (depending on the value of $\rho_T$), the EW hierarchy problem. The first setup is a realization of the RS model where the FOPT is at the TeV scale or above. The second setup is a simple RS-model extension with the FOPT below the TeV scale. In both setups, the spacetime is five dimensional (5D) and the metric has line element 
\begin{equation}
ds^2 = e^{-2A(z)} \eta_{MN} dx^M dx^N \,,
\label{eq:metric}
\end{equation}
where $A(z)$ is the warped factor and $x^M=(x^\mu,z)$ are conformal coordinates
with $M=0, \dots, 4$ and $\mu=0, \dots, 3$.

\subsection{High-energy setup: $\rho\gtrsim $ TeV}
We assume the spacetime given by Eq.~\eqref{eq:metric} has two boundaries $\mathcal B_a$ ($a=0,1$) along the fifth dimension: the boundary $\mathcal B_{0}$ is the ultraviolet (UV) brane and is located at $z=z_{0}=1/k$, where $k$ is a bulk parameter with mass dimension that fixes the AdS curvature between the two branes; the boundary $\mathcal B_{1}$ is the infrared (IR) brane and is placed at $z=z_1\equiv 1/\rho$. The parameter $\rho$ sets the energy scale of the IR brane and is hierarchically smaller than the energy scale of the UV brane due to the warp factor. Therefore, we can localize the SM at the IR brane and the EW hierarchy problem is solved if $\rho \sim 1\,\TeV$ (or alleviated if $\rho \gg 1\,\TeV$).

To stabilize the brane distance, $\log(z_1/z_0)$, we must break the conformal symmetry. We do it by introducing a (stabilizing) scalar field  $\phi$ that propagates in the bulk, i.e.~the spacetime between the branes. This field has the bulk potential $V(\phi)$ and brane potentials  %
\be
\Lambda_a(\phi)=\Lambda_a+\frac{1}{2}\gamma_a(\phi-v_a)^2\;.
\label{eq:brane_pot}
\ee
The brane potentials fix the values of $\phi$ at the branes, enforcing $\phi(z_a)=v_a$ in the stiff limit $\gamma_a\to \infty$. The bulk potential must be chosen wisely as it must satisfy the second-order 5D Einstein equations where the metric solution must behave as AdS$_5$ in the UV (i.e.~$A(z) \sim \log z$ at small $z$). Indeed, with such a behavior, the metric naturally induces a large hierarchy between the UV and IR energy scales, solving the EW hierarchy problem.

We obtain both $V(\phi)$ and $A(z)$ by solving the 5D Einstein equations with the so-called superpotential method~\cite{DeWolfe:1999cp}. We consider the superpotential $W(\phi)= 6kM_5^3 + u k \phi^2$, where $u>0$ is a dimensionless parameter controlling the backreaction of the scalar field on the 5D metric,  while $M_5$ is the 5D Planck scale. We also fix the brane tensions $\Lambda_a$ in Eq.~\eqref{eq:brane_pot} to $\Lambda_{0}=-W(v_{0})$ and $\Lambda_1=-W(v_1)+12 k M_5^3\lambda_1$, with $\lambda_1<0$ being a free detuning parameter~\cite{Goldberger:1999uk}. Such a superpotential and boundary conditions are known to yield the right AdS${}_5$ UV behavior (i.e.~$A(z)\sim \log z$ in the region $z\sim z_0$) in the case where the constant $u$ fulfills the condition $u\ll 1$, which means small backreaction~\cite{DeWolfe:1999cp}.

By breaking the conformal symmetry, and evaluating the classical action at the background of the stabilizing field $\phi$, an effective potential $V_{{\rm eff}}$ on the radion field $\chi(x)$ is induced, where $\chi(x)\equiv \rho+\tilde\chi(x)$ is the full radion field containing the degree of freedom describing the excitation $\tilde\chi(x)=\chi(x)-\rho$ around the location of the $\mathcal B_1$ brane. At the minimum of the potential, the radion acquires the vacuum expectation value $\langle\chi\rangle=\rho$, thus fixing the brane distance by $z_1/z_0=k/\rho$, fixed by the condition $v_1=v_0(z_1/z_0)^u$. 
Moreover, the second derivative of the  potential at the minimum permits to estimate the radion mass $m_\chi$. In the small-backreaction regime, such a mass can be approximated as~\cite{Megias:2023kiy}  
\be
m_{\chi}\equiv \hat m_{\chi} \rho,\quad \textrm{with}\quad \hat m_{\chi}\simeq \frac{\bar v_1}{15}u \log(k/\rho)   \,,
\label{eq:radion_mass}
\ee
where $\bar v_a = v_a /\sqrt{2 M_5^3}$ are two parameters of $\mathcal O(1)$ based on naturalness arguments.

On the other hand, at finite temperature, a second metric background is also compatible with the 5D Einstein equations when $A(z) \sim \log z$ at small $z$~\cite{Creminelli:2001th}. This solution is the 5D Anti-de-Sitter Schwarzschild (AdS-S) metric which describes the spacetime of a black hole (BH) with event horizon at $z=z_h$. As was the case for the AdS$_5$ solution, this metric background is symmetric under conformal invariance, although the latter is explicitly broken by the stabilizing field $\phi$, mainly via the brane potentials which fix the value of $\phi$ at the location of the different branes. This second solution is characterized by SU(N) strongly-coupled (massless) gauge fields that are deconfined. The AdS/CFT correspondence relates $N$ to $k$ through the equality

\be
N^2= 16 \pi^2  \left(\frac{M_5}{k} \right)^3 \quad \Longrightarrow \quad \frac{k}{M_P}=\frac{2\sqrt{2}\pi}{N}  \,,
\label{eq:kN}
\ee
where $M_P=2.4 \times 10^{15}$ TeV is the 4D reduced Planck scale, as $2M_5^3=k M_P^2$. Requiring the 5D gravity theory to be perturbative imposes the bound $N\gtrsim 5$~\cite{Agashe:2007zd,Chen:2014oha}. 

The fact that both the AdS-S and warped metric solutions coexist implies the existence of two spacetime background phases.  In the regime of small backreaction $u\ll 1$ and tiny temperature, the energy gap between the two phases is dominated by the difference  $E_0=\left|V_{{\rm eff}}(0) -  V_{{\rm eff}}(\rho)\right|$ which can be approximated as~\cite{Megias:2020vek}
\be
E_{0}\simeq \frac{3N^2\rho^4}{8\pi^2} |\lambda_1|  \,.
\label{eq:E0lambda1}
\ee   
Moreover, the free energy associated to these two phases presents a barrier between them, implying the existence  of a phase transition between the AdS and RS phases~\cite{Creminelli:2001th}. In the dual language, such a phase transition corresponds to a deconfined-confined FOPT.

On top of the above fields, also excitations of the graviton field $h_{\mu\nu}$ propagate in the bulk. The zero mode is massless and with flat profile (the 4D graviton) while the KK modes have masses $m_h\propto \rho$ and profiles localized toward the $\mathcal B_1$ brane.


\subsection{Low-energy setup: $\rho\lesssim $ TeV}
In the previous setup, the SM is localized at the IR brane. This is not a viable option if the IR energy scale, $\rho$, is much below the EW scale. At the same time, localizing the SM at the UV brane is not a good alternative neither, as it does not solve the hierarchy problem at all. We tackle this issue by introducing a third (SM) brane  $\mathcal B_T$ located at $z=z_T=1/\rho_T$, with $z_0 < z_T < z_1=1/\rho$, with the goal of achieving $\rho_T\gtrsim 1\,\TeV$ and $\rho_T \gg \rho$, and localizing the SM there~\cite{Lykken:1999nb,Hatanaka:1999ac,Kogan:1999wc,Gregory:2000jc,Kogan:2000xc,Agashe:2016rle,Agashe:2016kfr,Lee:2021wau,Girmohanta:2023sjv}. With this goal in mind, we dub $\mathcal B_T$ as TeV brane. 

Stabilizing the positions of the TeV and IR branes requires breaking the conformal invariance. We do it by repeating the rationale of the previous setup for both the TeV and the IR brane. 
We then introduce the superpotentials $W(\phi) = 6 k_1 M_5^2+ u_1 k_1 \phi^2$ in Region 1,  and $W(\phi) = 6 k_2 M_5^2+ u_2 k_2 \phi^2$ in Region 2, where Region 1 and Region 2 are the two bulk regions between the branes, $z_0\leq z\leq z_T$ and $z_T\leq z\leq z_1$, respectively. On the branes, the stabilizing scalar field has brane potentials $\Lambda_a(\phi)$ analogous to those in Eq.~\eqref{eq:brane_pot}, with $a=0,T,1$.

In Region 1, the curvature is associated with the parameter $k_1$, related to the number of degrees of freedom $N_1$ of the sector described by the $\mathcal B_T$ brane.  The location of the brane $\mathcal B_T$ with respect to the $\mathcal B_0$ brane is stabilized as  $v_0=v_T(z_0/z_T)^{u_1}$  by a \emph{heavy} radion $\chi_1(x)$ (describing excitations $\tilde\chi_1(x)\equiv \chi_1(x)-\rho_T$ around the location of the $\mathcal B_T$ brane, and with a mass ($m_{\chi_1}\sim u_1 \rho_T$) parametrically controlled by the parameter $\rho_T$), which acquires vacuum expectation value  $\langle\chi_1 \rangle = \rho_T$. This radion is the zero mode of a 5D field with a bulk profile localized toward the $\mathcal B_T$ brane.

In Region 2, the curvature and number of degrees of freedom of the dual theory associated with the $\mathcal B_1$ brane are given by $k_2$ and $N_2$. The location of the brane is stabilized by the potential of the \emph{light} radion $\chi_2(x)$ (which similarly describes excitations $\tilde\chi_2(x)\equiv \chi_2(x)-\rho$ around the location of the $\mathcal B_1$ brane, with a mass ($m_{\chi_2}\sim u_2\rho$) parametrically controlled by the parameter $\rho$), and is given by $v_T=v_1(z_T/z_1)^{u_2}$. The vacuum expectation value of this radion is $\langle \chi_2 \rangle=\rho$. The 5D profile of the radion $\chi_2$ is localized toward the $\mathcal B_1$ brane.

The setup thus exhibits two radion FOPTs, one at the scale $\rho_T\gtrsim $\,TeV and one at the scale $\rho \lesssim 1\,$TeV. In this paper, we are only interested in the latter FOPT whenever we consider this setup.\footnote{We postpone the full analysis of both phase transitions in the general parameter setup for future work.} Therefore, for simplicity, we assume the particular case $k_1\simeq k_2=k$, (i.e.~$N_1\simeq N_2=N$) and $u_1\simeq u_2=u$. In this case, the heavy radion $\chi_1$ is very massive (as $m_{\chi_1}^2\propto \frac{1}{k_1-k_2}$)~\cite{Lee:2021wau}, and decouples from the low-energy theory. On the contrary, the light radion $\chi_2\equiv \chi$, which stabilizes the $\mathcal B_1$ brane by acquiring the vacuum expectation value $\rho$, behaves like the radion of the previous setup. The location of the two branes are fixed by the dynamical relations $v_T=v_1(z_T/z_1)^u$ and $v_0=v_1(z_0/z_1)^u$. Setting $\bar v_0 < \bar v_T < \bar v_1$ with values of $\mathcal{O}(1)$ then naturally lead to the huge hierarchy among the Planck, TeV and IR scales, i.e.~$\rho_T \gtrsim 1\,\TeV$ and $\rho \lesssim 1\,$TeV.

On top of the above fields, the excitations of the graviton field $h_{\mu\nu}$ propagate in the bulk as in the high-energy setup, with a massless zero mode and KK modes with masses $m_h\propto \rho$ and profiles localized toward the $\mathcal B_1$ brane. Their coupling with the SM fields is then further suppressed by the distance between the $\mathcal B_1$ and $\mathcal B_T$ branes $z_T/z_1$~\cite{Lee:2021wau}.


\subsection{Setups summary}
We summarize these setups in Tab.~\ref{tab:setups}. In presenting our analysis, we implicitly swap from  the high-energy setup to the low-energy setup  whenever we move from $\rho>1\,\TeV$ to $\rho<1\,\TeV$, and vice versa.

\begin{table}[htb]
  \centering
  \begin{tabular}{|l||c|c|c||c|c|}
    \hline
    \multicolumn{6}{|c|}{\bf Model setups} \\
    \hline
    \hline
 {\bf IR scale}    & \multicolumn{3}{|c||}{$\rho \lesssim 1 \, \TeV $}  & \multicolumn{2}{|c|}{$\rho \gtrsim 1 \, \TeV$}    \\
    \hline
 {\bf Branes}  &   $\mathcal B_0$  \hspace{0.cm}  &  $\mathcal B_T$  & $\mathcal B_1$    &  $\hspace{0.0cm} $ $\mathcal B_0$ \hspace{0.0cm}  & $$ $\mathcal B_1$   $$     \\
    \hline
{\bf Brane positions}   &   \multicolumn{3}{|c||}{$z_0 < z_T < z_1$}     &  \multicolumn{2}{|c|}{$z_0 < z_1$}     \\ 
   \hline
 {\bf Scales}   & \multicolumn{3}{|c||}{$k > \rho_T > \rho$}   & \multicolumn{2}{|c|}{$k > \rho$}  \\
    \hline
 {\bf SM localization}  &                & SM  &            &  & SM   \\
     \hline 
  \end{tabular}
\caption{\it Model setups. The two possibilities: $\rho \lesssim 1$ TeV and $\rho \gtrsim 1$ TeV are summarized. The localization of the SM is indicated for each setup. It is assumed that $\rho_T \simeq 1 \, \TeV$.}
\label{tab:setups}
\end{table}

The two setups exhibit a similar phenomenology once, in the low-energy setup with the hierarchy $z_1/z_T\gg1$, the heavy radion is decoupled and the heavy-radion dynamics does not impact the light-radion FOPT. Indeed, the KK excitations only appear for the fields propagating in the bulk and, since the SM particles are bound in a brane, only the bulk field $\phi$ and the metric provide 5D resonances. However, $\phi$ does not provide an independent degree of freedom, so that the gravitons and the radion of the IR brane drive the BSM signatures of the model. Their phenomenology depends on the following free parameters: $N$, $\rho$, $\lambda_1$, $u$ and $v_a$. As Eqs.~\eqref{eq:radion_mass} and~\eqref{eq:E0lambda1} show, $N$, $\rho$ and $\lambda_1$ control the energy gap $E_0$, which the FOPT 
duration parameter $H_*/\beta $ depends on, while $u$ and $v_a$ control the (dimensionless) IR-brane radion mass $\hat m_\chi$. Our analysis can then use $N$, $\rho$, $\beta/H_*$ and $\hat m_\chi$ as proxies of the aforementioned free parameters.

\section{Strong first-order phase transition}
\label{sec:PhaseTransition}
\noindent 
In the 5D theory there are two vacua: the usual RS vacuum where the free energy is provided by the radion potential $V(\chi)$ fixing the inter-brane distance at $\langle\chi\rangle= \rho$, and the AdS-S vacuum which corresponds to a thermal state with the Hawking temperature $T_h$. In the holographic picture these two vacua correspond respectively to the confined and deconfined phases.

The free energies in these phases, $F_c$ and $F_d$, are given by
\be
F_c(T) = -E_0 - \frac{\pi^2}{90}g_c^{\rm eff}T^4 \,, \qquad  F_d(T) = -\frac{\pi^2}{8}N^2 T^4-\frac{\pi^2}{90}g_d^{\rm eff}T^4  \,, 
\label{eq:F}
\ee
where $E_0=V(0)-V(\rho)>0$ is the $T=0$ potential gap between the two phases, and $g_c^{\rm eff}$ ($g_d^{\rm eff}$) is the effective number of degrees of freedom in the confined (deconfined) phase, which we assume to be $g_c^{\rm eff}\simeq g_d^{\rm eff}\simeq g_{\rm SM}= 106.75$, corresponding to the SM number of degrees of freedom. 

The phase transition can start at temperatures $T<T_c$ where the critical temperature, $T_c$, is defined as $F_c(T_c)=F_d(T_c)$, which leads from Eq.~(\ref{eq:F}) to
\be
E_0=\frac{\pi^2}{8}N^2 T_c^4  \,,
\label{eq:Tc}
\ee
where the terms in $g_c^{\rm eff}$ and $g_d^{\rm
  eff}$ cancel out due to our assumption $g_c^{\rm eff}\simeq g_d^{\rm eff}$~\footnote{Even in the cases
where the terms in $g_c^{\rm eff}$ and $g_d^{\rm eff}$ do not cancel,
$\frac{\pi^2}{90}(g_c^{\rm eff}-g_d^{\rm eff})\ll \frac{\pi^2}{8}N^2$,
for $N>10$, so that these terms can be omitted from the balance
equation (\ref{eq:Tc}).}. 
Using now the expression for $E_0$ in Eq.~\eqref{eq:E0lambda1} we obtain 
\be
T_c^4=\frac{3|\lambda_1|\rho^4}{\pi^4}  \,.  \label{eq:Tclambda1}
\ee

Since both phases are separated by a
potential barrier, the phase transition is of first order and proceeds by bubble nucleation
of the confined (broken) phase in the deconfined (symmetric) phase. While at high $T$ this
process is driven by thermal fluctuations with $O(3)$ symmetric
euclidean action, at low $T$ the process is driven by quantum
fluctuations with $O(4)$ symmetric euclidean action. The latter then holds in the regime of large supercooling, which is the case we are interested in this paper~\footnote{We have checked by numerical calculations that the phase transition is driven in all cases by the $O(4)$ symmetric solutions, and not by thermal fluctuations with symmetry $O(3)$.}.

The bounce action driven by fluctuations with $O(4)$ symmetry reads
\be
S_4=2\pi^2\int d\sigma \sigma^3\frac{3N^2}{4\pi^2}\left[\frac{1}{2}\left(\frac{\partial\chi}{\partial\sigma} \right)^2+V(\chi,T) \right]  \,,
\ee
where $\sigma=\sqrt{\vec x^2+\tau^2}$, with $\tau$ being the euclidean time. The corresponding equation of motion is given by
\be
\frac{d^2\chi}{d^2\sigma}+\frac{3}{\sigma}\frac{d\chi}{d\sigma}-\frac{\partial V}{\partial\chi}=0   \,,
\label{eq:EoM}
\ee
with boundary conditions $\chi(0)=\chi_0$ and $d\chi/d\sigma=0$. The corresponding temperature is obtained by equating the kinetic energy of the radion at the origin,
$\chi=0$, with the thermal energy, so that the barrier between the two vacua can be overcome~\footnote{This approach is an excellent approximation of the full bounce solution, including both the radion and the degree of freedom associated to the position of the AdS-Swarzschild horizon. For details see e.g.~Ref.~\cite{Megias:2020vek}.}.

By definition, the nucleation temperature $T_n$ is reached when there is one bubble produced per causal Hubble volume. At this temperature, the action has value  $S_4(T_n)=S_n$, where $S_n$ is computed in App.~\ref{sec:AppA}. To determine $T_n$, we then solve numerically Eq.~(\ref{eq:EoM}) and check that the condition $S_4(T_n)=S_n$ is met. In App.~\ref{sec:AppB} we compare this approach with the thick-wall approximation, which is supposed to be reliable for supercooled FOPT. The comparison shows that the thick-wall approximation does not describe accurately the numerical results in all our benchmark scenarios, so that, in order to simplify the numerical analysis, we implement semi-analytical (thick-wall inspired) formul$\ae$ for the different temperatures that fit the numerical result in all the considered cases. 

\begin{figure}[tb]
  \centering
    \includegraphics[width=4.6cm]{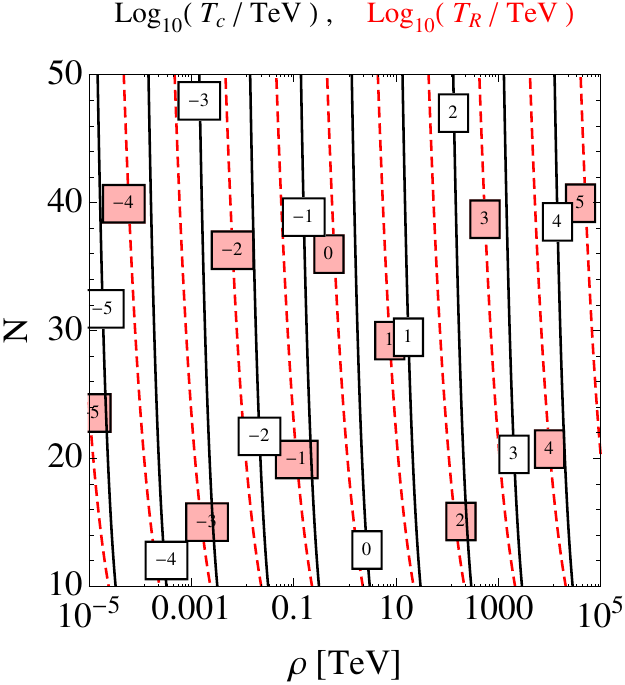} \hspace{0.2cm}   \includegraphics[width=4.6cm]{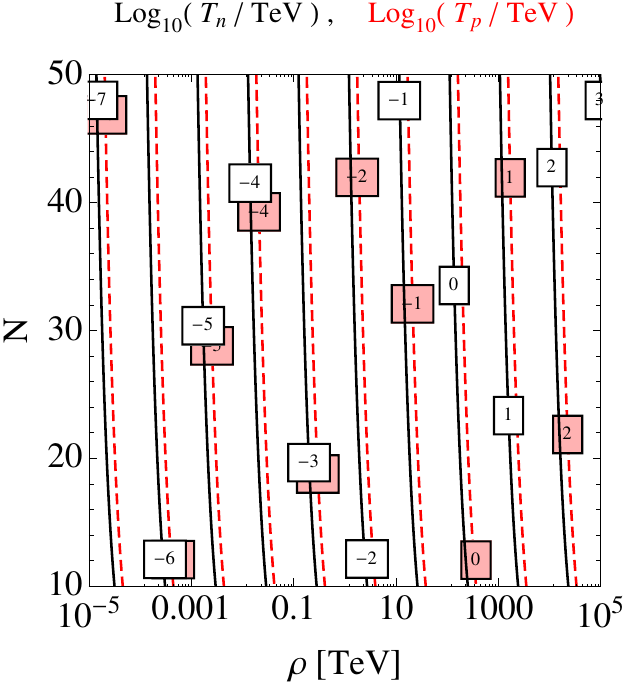} \hspace{0.2cm}   \includegraphics[width=4.7cm]{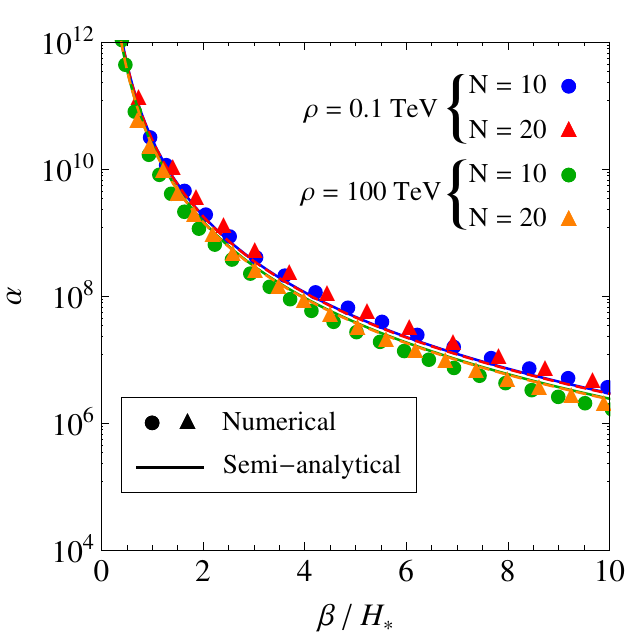}    \\ 
  \caption{\it \textbf{Left and middle panels:}  Contour plots of $\log_{10}(T_c / \textrm{TeV})$ and $\log_{10}(T_R / \textrm{TeV})$ (left panel), and $\log_{10}(T_n / \textrm{TeV})$ and $\log_{10}(T_p / \textrm{TeV})$ (middle panel) in the plane $\{\rho,N\}$, obtained from the semi-analytical approximations of Eqs.~(\ref{eq:Tcsection})-(\ref{eq:TpTnsection}). $T_c$ and $T_n$ are displayed in solid, black lines, while $T_R$ and $T_p$ are shown in dashed, red lines. The value $\beta/H_\ast = 6$ is assumed. \textbf{Right panel:} Plot of $\alpha \equiv \alpha(T_n)$ as a function of $\beta/H_\ast$ for $\alpha$  computed from the numerical solution of the bounce equation (dots) and from the semi-analytical expressions (solid lines).}
\label{fig:temperatures}
\end{figure}

As anticipated in the previous section, we can trade the free parameter $\lambda_1$ by the parameter $\beta/H_*$ that characterizes the tunneling rate: 
\be
\beta/H_\ast =\left.T\frac{dS_4}{dT}\right|_{T=T_n}.
\label{eq:betaH}
\ee
In our thick-wall inspired expressions, we limit ourselves to $\beta/H_\ast\lesssim 10$, for which the formation of PBH is relevant, as we will see in the next section~\footnote{For moderately strong FOPT, we do not need to distinguish between the Hubble parameter at the nucleation temperature~$H_n$, and the Hubble parameter during percolation $H_\ast$, as their values are very similar~\cite{Caprini:2019egz}. In fact, in our model and for the considered values of $\beta/H_*$ we can see from the middle panel of Fig.~\ref{fig:temperatures} that both temperatures are very close to each other.}.  As for the energy balance, we quantify it by the parameter $\alpha(T_n)$ given by
\be
\alpha(T)=\frac{|F_d(T)-F_c(T)|}{\rho_d(T)}=\frac{1}{3}\left\{\left( \frac{T_c}{T}\right)^4-1\right\}\,,  \label{eq:alpha}
\ee
providing $\alpha\geq 0$ for $T\leq T_c$. Supercooled phase transitions typically lead to $\alpha(T_n)\gg 1$.

The semi-analytical values of the different temperatures, critical $T_c$, nucleation $T_n$, reheating $T_R$ and percolation $T_p$, are then given by the following expressions (see App.~\ref{sec:AppB} for details) 
\begin{align}
T_c &=  a_c \sqrt{\frac{3N}{S_n}} \left(4S_n+\frac{\beta}{H_\ast} \right)^{1/4}\frac{\rho}{2\pi} \,,\qquad a_c\simeq 0.9 \,,  \label{eq:Tcsection} \\
T_n &= a_n \sqrt{\frac{3N}{S_n}} \left(\frac{\beta}{H_\ast} \right)^{b_n} \frac{\rho}{2\pi}  \,,\qquad  a_n\simeq 8.25 \times 10^{-3} \,,\ b_n = 1 \,,  \label{eq:Tnsection} \\ 
T_R&\simeq a_R \frac{3^{3/4}5^{1/4}}{2\sqrt{2}\pi} \frac{ \left(4S_n+\frac{\beta}{H_\ast} \right)^{1/4}}{(g_c^{\rm eff})^{1/4}\sqrt{S_n}} \, N \rho  \,,   \qquad a_R\simeq 0.9 \,, \label{eq:TRsection}  \\
\frac{T_p}{T_n} &=\left\{ 
\begin{array}{lc}
c_1 \left(\frac{\beta}{H_\ast}\right)^{1/8}\,,\qquad c_1\simeq 0.548 &\quad \textrm{for}\quad\frac{\beta}{H_\ast}\lesssim 8 \\
d_0+d_1\left(\frac{\beta}{H_\ast}\right)\,, \quad d_0\simeq 0.671,\ d_1\simeq 5.1\times 10^{-3}&\quad\textrm{for}\quad  \frac{\beta}{H_\ast}\gtrsim 8
\end{array}
\right.    \,.   \label{eq:TpTnsection}
\end{align}
From Eq.~(\ref{eq:Tclambda1}) and Eq.~(\ref{eq:Tcsection})  we explicitly see how one can trade the parameter $\lambda_1$ by $\beta/H_\ast$ as
\be
\frac{\beta}{H_\ast}=4S_n\left( \frac{4}{3} \frac{S_n}{a_c^4N^2}|\lambda_1|-1 \right)  \,,
\label{eq:betaH_Sn_lambda1}
\ee
which gives a mild lower bound on $|\lambda_1|$: from $\beta/H_\ast\gtrsim 1$ and $S_n\gg 1$, one obtains
\be
|\lambda_1|\geq  \frac{3}{4} \frac{a_c^4 N^2}{S_n}  \,.
\ee
We have tested that such expressions are practically independent of the radion mass $\hat m_\chi$, and hence of $v_a$ and $u$, at least in the parameter region we are interested in, i.e.~$\beta/H_\ast \lesssim 10$.

%

In Fig.~\ref{fig:temperatures} (left and middle panels) we show contour lines of the temperatures relevant for the phase transition $T_c$, $T_n$, $T_R$ and $T_p$ in the plane $\{\rho,N\}$. These temperatures mostly depend on $\rho$, and they just have a mild dependence on~$N$. In the right panel of this figure, we also display the behavior of $\alpha \equiv \alpha(T_n)$ as a function of $\beta/H_\ast$ for different values of $\rho$ and $N$ in a regime in which the numerical computation of the bounce equation was possible, and compare these results with the semi-analytical approximation of $\alpha$ given by Eq.~(\ref{eq:alpha}) together with the semi-analytical formul$\ae$  of $T_c$ and $T_n$, Eqs.~(\ref{eq:Tcsection})-(\ref{eq:Tnsection}).  As one can notice, $\alpha$ is extremely large. 

In Table~\ref{tab:benchmark} we define eight
benchmark points (top block) and quote the values of the FOPT semi-analytic quantities obtained at such points (next-to-top block). The input for $\hat m_\chi$ is omitted as it plays a negligible role in the FOPT phenomenology. 
The table also displays the numerical outputs that will be obtained in the following sections. We defer to those sections the definitions and  discussions of these outputs.

\begin{table}[htb]
  \centering
  \resizebox{15cm}{!}{
  \begin{tabular}{||c|c|c|c|c|c|c|c|c||}
    \hline
    \multicolumn{9}{|c|}{ {\bf Benchmark points} } \\
    \hline
         \hline
{\bf Bench. point}          & \textbf{P}${}_{\mathbf 1}$  &  \textbf{P}${}_{\mathbf 2}$   &  \textbf{P}${}_{\mathbf 3}$    &  \textbf{P}${}_{\mathbf 4}$  &  \textbf{P}${}_{\mathbf 5}$  &  \textbf{P}${}_{\mathbf 6}$    &  \textbf{P}${}_{\mathbf 7}$    &  \textbf{P}${}_{\mathbf 8}$     \\
     \hline
         \hspace*{1cm}$\boldsymbol{N} \hspace{1.1cm}$  & 10  & 50 & 10 & 50 & 10 & 50 & 10 & 50    \\
     \hline     \hspace*{1cm}$\boldsymbol{\rho}~ \textbf{[TeV]} \hspace{1.1cm}$  & $5 \cdot 10^{-4}$  & $5 \cdot 10^{-4}$ & 1 & 1 & 10 & 10 & 100 & 100    \\
               \hline
     $\boldsymbol{-\lambda_1}$  & 0.284 & 7.38 &0.346 &9.08 &0.371 &9.74 &0.398 &10.52   \\
     \hline\hline    
         $T_c \; [\TeV]$      &  $1.5 \cdot 10^{-4}$   & $3.5 \cdot 10^{-4}$   &  $0.32$  & $0.73$   & $3.3$   &  $7.4$   & $33$   &  $75$      \\
     \hline
        $T_n \; [\TeV]$      &  $1.5 \cdot 10^{-6}$    & $3.7 \cdot 10^{-6}$   & $4.1 \cdot 10^{-3}$   &  $9.6 \cdot 10^{-3}$   & $4.2 \cdot 10^{-2}$    &  $9.2 \cdot 10^{-2}$   &  0.42     &  $0.98$      \\
     \hline
         $T_p\; [\TeV]$      & $1.1 \cdot 10^{-6}$     & $2.5 \cdot 10^{-6}$   & $2.8 \cdot 10^{-3}$   &  $6.7 \cdot 10^{-3}$   &  $2.9 \cdot 10^{-2}$    &  $6.4 \cdot 10^{-2}$   &   $0.29$  &  $0.68$      \\
     \hline
         $T_R\; [\TeV]$      & $2.1 \cdot 10^{-4}$     & $1.1 \cdot 10^{-3}$   & $0.44$   & $2.2$   & $4.5$    & $23$   & $46$   & $230$       \\
     \hline
         $\alpha$      &  $3.2 \cdot 10^7$   & $2.6 \cdot 10^7$    & $1.3 \cdot 10^7$     &  $1.1 \cdot 10^7$   &  $1.3 \cdot 10^7$   &  $1.4 \cdot 10^7$    &  $1.3 \cdot 10^7$    &  $1.2 \cdot 10^7$      \\
     \hline
         $\left( \beta/H_\ast  \right)_{\textrm{min}}$     & $5.7$    & $5.9$   &  $6.8$  & $7.0$   & $6.7$   &  $6.5$   &  $6.5$   &  $6.6$      \\
     \hline\hline
         $M_{\PBH}/M_\odot$      &  $0.21$    & $8.3 \cdot 10^{-3}$   &  $4.8 \cdot 10^{-8}$   &  $1.9 \cdot 10^{-9}$   & $4.6 \cdot 10^{-10}$   & $1.8 \cdot 10^{-11}$   &  $4.5 \cdot 10^{-12}$  &  $1.7 \cdot 10^{-13}$      \\
     \hline
     $f_\PBH$      &  $0.066$   & $0.035$   & $2.2 \cdot 10^{-4}$   & $6.4 \cdot 10^{-5}$   &  $8.7 \cdot 10^{-3}$   & $1$    & $1$   &  $1$     \\
       \hline
 Tuning       $\Delta$  &  $6.9 \cdot 10^3$   & $7.2 \cdot 10^3$   & $8.6 \cdot 10^3$   & $8.8 \cdot 10^3$   &  $7.8 \cdot 10^3$   & $6.8 \cdot 10^3$    & $6.8 \cdot 10^3$   &  $6.7 \cdot 10^3$     \\     
 \hline    
 BH spin  $S_{\PBH}$     & $1.6 \cdot 10^{-3}$    &    $1.5 \cdot 10^{-3}$   &  $1.0 \cdot 10^{-3}$    &  $0.93 \cdot 10^{-3}$     &  $1.0 \cdot 10^{-3}$    & $1.1 \cdot 10^{-3}$     &  $1.1 \cdot 10^{-3}$    & $1.1 \cdot 10^{-3}$         \\
     \hline\hline
         $h^2 \bar\Omega_{\GW}$      & $2.5 \cdot 10^{-8}$    & $2.3 \cdot 10^{-8}$    & $1.7 \cdot 10^{-8}$    &  $1.6 \cdot 10^{-8}$   & $1.8 \cdot 10^{-8}$    & $1.9 \cdot 10^{-8}$    &  $1.9 \cdot 10^{-8}$   &  $1.8 \cdot 10^{-8}$      \\
     \hline
         $f_p \; [\textrm{Hz}]$      & $2.2 \cdot 10^{-8}$    &  $1.1 \cdot 10^{-7}$  & $5.5 \cdot 10^{-5}$   & $2.9 \cdot 10^{-4}$   &  $5.5 \cdot 10^{-4}$   & $2.7 \cdot 10^{-3}$   &  $5.5 \cdot 10^{-3}$   &  $2.8 \cdot 10^{-2}$      \\
     \hline 
  \end{tabular} \label{tab:BPs}
  }
\caption{\it Benchmark points for the FOPT phenomenology analysis. In the top block, the input values. In the next-to-top block, the outputs about the FOPT quantities detailed in Sec.~\ref{sec:PhaseTransition}. In the third block, the outputs concerning the PBH quantities defined in Sec.~\ref{sec:PBH}. In the bottom block, the outputs on the SGWB predictions discussed in Sec.~\ref{sec:GW}.  For a given set of $\{N,\rho\}$ input, the value of $\lambda_1$ is chosen to maximize the PBH abundance experimentally allowed, as explained in Sec.~\ref{sec:PBH}.}
\label{tab:benchmark}
\end{table}

\section{Primordial black hole formation}
\label{sec:PBH}
\noindent
A FOPT can lead to formation of PBHs
in different ways~\cite{Liu:2021svg,Kawana:2022olo,Gouttenoire:2023naa,Lewicki:2023ioy}.  In particular, due to the stochastic nature of
nucleation, there is a probability that some regions remain for a
longer time in the false vacuum while the space around them gets
filled by true vacuum.  Reheating in the true vacuum leads to
radiation, so that its energy density $\rho_{\rm true}$ dilutes, while
the energy density in the false vacuum is constant $\rho_{\text{false}} \simeq
V_\Lambda$ if the FOPT transition is extremely supercooled.  Thus regions in the false vacuum become relatively denser
as quantified by $\delta \equiv \rho_{\text{false}} /
\rho_{\text{true}} - 1$, with the effect being more pronounced if the time gap between nucleation and percolation is long, i.e.~the FOPT is slow or, equivalently, $\beta/H $ is small.  If a region in the false vacuum
(approximated as roughly-spherical) has radius larger than roughly the
inverse Hubble parameter, it forms a PBH when the density
contrast exceeds the critical value $\delta_{\text{c}} \approx 0.45$
\cite{2002.12778}. This means that PBHs can form due to the
collapse of overdense regions during a supercooled FOPT~\cite{Gouttenoire:2023naa,Conaci:2024tlc}.

In the limit of $\alpha\gg 1$, the probability that a Hubble patch collapses into a PBH, $\mathcal P_{\rm coll}$, can be approximated by the analytic formula~\cite{Gouttenoire:2023naa,Conaci:2024tlc}
\be
  \mathcal P_{\textrm{coll}}  \simeq \exp\left[ - a \left( \frac{\beta}{H_\ast} \right)^b (1 + \delta_c)^{c \frac{\beta}{H_\ast}} \right] \,,   \label{eq:Pcoll} 
\ee
where $a\simeq 0.5646$, $b\simeq 1.266$, $c\simeq 0.6639$ and $\delta_c\simeq 0.45$. As we can see, the collapse probability does not really depend on the scale of the phase transition but on the parameter $\beta/H_\ast$, so that the probability becomes exceedingly small for large values of the parameter $\beta/H_\ast$.

In first approximation, PBHs form with a mass roughly given by the mass within the sound horizon volume, $c_sH^{-1}$,  at bubble collision time, leading to
\be
M_{\PBH} \approx 3.7 \times 10^{-8} M_{\odot} \left(\frac{ 106.75}{g_{c}^{\rm eff}(T_R)}\right)^{1/2} \left(\frac{0.5\, \TeV}{T_R}\right)^{2}  \,,
\label{eq:MPBH}
\ee
where $M_{\odot} \simeq 2\times 10^{30}\,\text{kg}$ is the solar mass~\cite{Gouttenoire:2023naa}.
Their abundance, expressed as the fraction of DM in the form of PBH today, is given by
\be
 f_{\PBH} =\frac{\rho_{\PBH,0}}{\rho_{\rm DM,0}}\simeq \left( \frac{\mathcal P_{\textrm{coll}}}{6.2 \times 10^{-12}} \right) \left( \frac{T_R}{0.5 \, \textrm{TeV}} \right)  \,.  \label{eq:fPBH}
\ee
Their spin can be estimated using the peak theory formalism and
is parameterized by the dimensionless Kerr parameter $a_\ast$~\cite{Banerjee:2023qya,Conaci:2024tlc,Arteaga:2024vde}:
\be
S_{\PBH} \equiv \langle a_\star^2\rangle^{1/2}\simeq 4.01\times 10^{-3}\frac{\sqrt{1-\gamma^2}}{1+0.036\left[ 2.7-2\log_{10}\left( \frac{f_{\PBH}}{10^{-7}}\right)-\log_{10}\left(\frac{M_{\PBH}}{M_{\odot}} \right) \right]}  \,,  \label{eq:spin_BH}
\ee
where we use the reference value $\gamma\simeq 0.96$.
  
\begin{figure}[th]
  \centering
  \includegraphics[width=5.1cm]{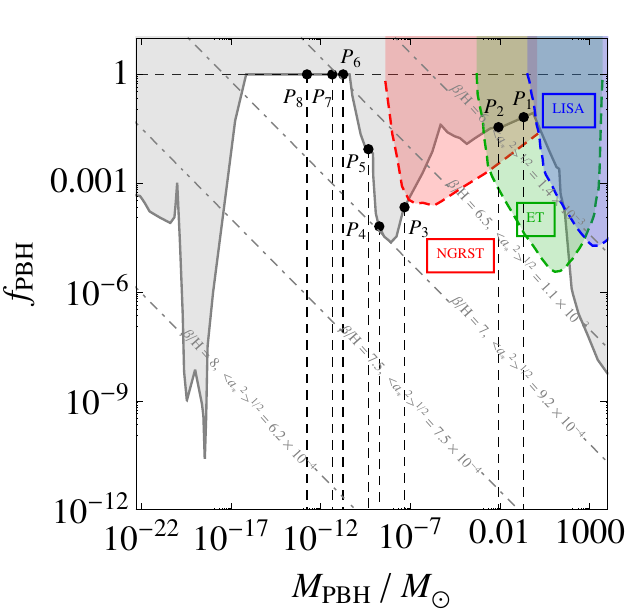}  \hspace{0.2cm}  
   \includegraphics[width=5.0cm]
  {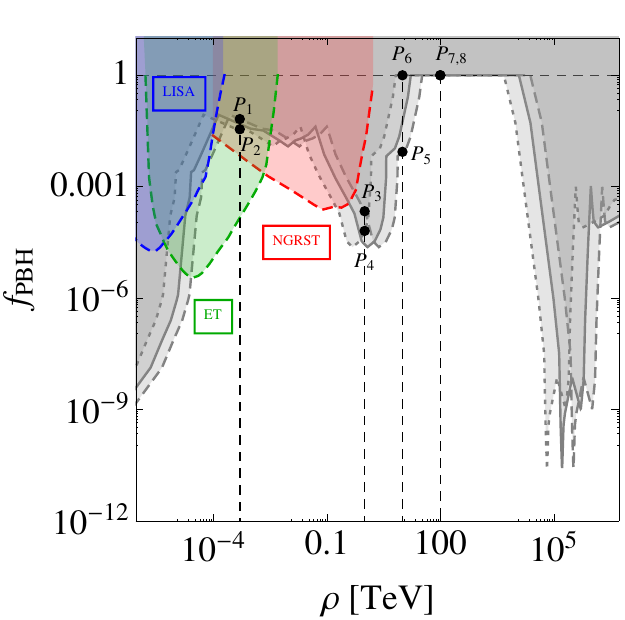} \hspace{0.2cm}   \includegraphics[width=4.4cm]{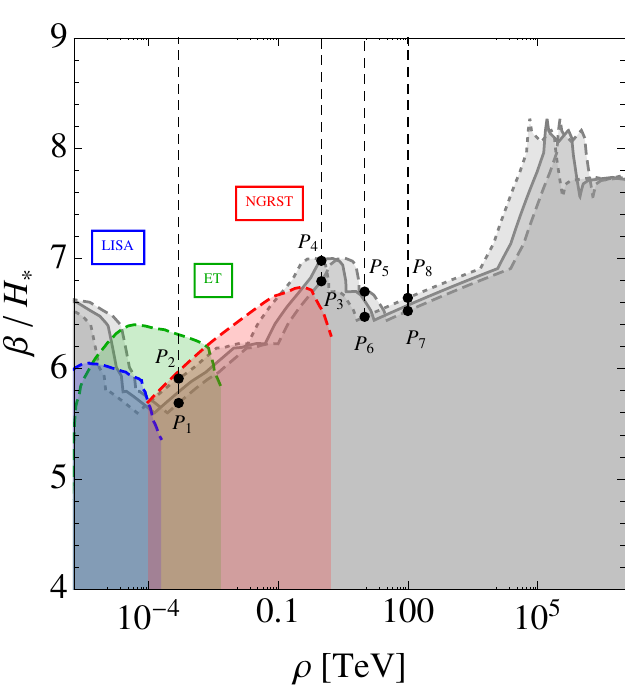} \\
  \caption{\it \textbf{Left panel:} 
    Excluded region (gray area) and  NGRST, ET and LISA parameter reach (red, blue and green areas) 
    in the plane $\{\rho,M_{\PBH}\}$. Bullet points correspond to the benchmark points of Table~\ref{tab:benchmark}. \textbf{Middle  panel:}  As for the left panel, but in the plane 
    $\{f_{\PBH}, \rho \})$ for $N=10, 20, 50$ (dashed, solid, dotted border). Parameter reaches shown only for  $N=20$.  \textbf{Right panel:} As for the central panel, but in the plane $\{\beta/H_\ast, \rho \}$.}
\label{fig:MPBHrho}
\end{figure}

In the left panel of Fig.~\ref{fig:MPBHrho} we summarize the PBH experimental bounds and future probes. The gray area corresponds to the parameter region $\{ f_{\rm PBH}, M_{\rm PBH}\}$ 
experimentally ruled out in scenarios where the PBH mass distribution is extremely peaked, say monochromatic, which is what we expect in our radion FOPT. The gray area includes the following bounds  (see Refs.~\cite{Green:2020jor,Saha:2021pqf,Laha:2019ssq,Ray:2021mxu} for details):
\begin{itemize}
\item Hawking evaporation of PBH is relevant at masses $5 \times 10^{-24}M_\odot\lesssim M_{\rm PBH}   \lesssim 10^{-16}M_\odot$ and  implies constraints using data from CMB\,\cite{Clark:2016nst}, EDGES\,\cite{Mittal:2021egv},  {\sc Integral}\,\cite{Laha:2020ivk,Berteaud:2022tws}, {\sc Voyager}\,\cite{Boudaud:2018hqb}, 511\;keV\,\cite{DeRocco:2019fjq} and EGRB\,\cite{Carr:2009jm}. 

\item {Micro-lensing} observations from HSC~\cite{Niikura:2017zjd}, 
EROS\,\cite{EROS-2:2006ryy}, {\sc Icarus}\,\cite{Oguri:2017ock}, including a hint for PBH in OGLE\,\cite{Niikura:2019kqi}, are relevant for masses $5 \times 10^{-11} M_\odot \lesssim M_{\rm PBH}   \lesssim 10^4 M_\odot$.

\item LVK detections binds the properties of the stellar-mass black population, imposing a constraint on PBHs with $1 \, M_{\odot} \lesssim M_{\rm PBH}\lesssim 100 \,M_{\odot}$~\cite{Franciolini:2022tfm,Kavanagh:2018ggo,Hall:2020daa,Wong:2020yig,Hutsi:2020sol,DeLuca:2021wjr,Franciolini:2021tla}. 

\item Heavy PBHs are expected to accrete, leading to signatures in the CMB. The null observation of this effect binds the region $ M_{\rm PBH} \gtrsim 10\, M_{\odot}$~\cite{Serpico:2020ehh,Piga:2022ysp}.
\end{itemize}
The colored areas instead correspond to the PBH parameter reach achivable at forthcoming experiments. 
Micro-lensing future sensitivity of the NGRST~\cite{DeRocco:2023gde} is depicted in red. ET and LISA sensitivities to BPH mergers are in green and blue, respectively~\cite{Pujolas:2021yaw} (see also Refs.~\cite{DeLuca:2021hde,Franciolini:2022htd,Martinelli:2022elq,Franciolini:2023opt,Marcoccia:2023khb, Branchesi:2023mws,Marriott-Best:2024anh}).
Therefore, the PBH mass window able to explain the entire DM abundance is
  \be
  7 \times 10^{-17}\,  M_\odot \lesssim M_{\text{PBH}} \lesssim 4\times10^{-11}\, M_\odot \,.
  \label{eq:PBHmassrange}
  \ee

By looking at Eqs.~\eqref{eq:Pcoll}, \eqref{eq:MPBH} and \eqref{eq:fPBH}, one can notice that $f_{\rm PBH}$ only depends on $\beta/H_\ast$ once $M_{\text{PBH}}$ (or equivalently, $T_R$) is fixed. 
Moreover,  $M_{\rm PBH}$ (or equivalently  $T_R$) is a function of $\rho$, $N$ and $\beta/H_*$ via  Eqs.~\eqref{eq:TRsection} and \eqref{eq:Snnumerical}. This allows us to map,   for a fixed $N$, 
every point of the PBH space parameter 
 $\{f_{\rm PBH} , M_{\rm PBH} \}$ to a point of the plane  $\{f_{\rm PBH} , \rho \}$ or  to the plane $\{\rho ,  \beta/H_\ast \}$. Using these mappings, in  Fig.~\ref{fig:MPBHrho}  we recast the aforementioned constraints and future sensitivities given in the plane $\{f_{\rm PBH} , M_{\rm PBH} \}$ (left panel) in terms of other parameters (middle and right panels) for $N=10, 20, 50$ (dashed, solid, dotted borders). 
As the central panel shows, in our (high-energy) 5D setup the PBH mass window in Eq.~\eqref{eq:PBHmassrange} corresponds to $\rho \in [33.2, 2.40 \times 10^{4}] \, \textrm{TeV}$ for  $N = 10$, $\rho \in [16.6, 1.20 \times 10^{4}] \, \textrm{TeV}$ for  $N = 20$, and $\rho \in [6.6, 4.8 \times 10^{3}] \, \textrm{TeV}$ for  $N = 50$, which arises for  $15 \, \TeV \lesssim T_R  \lesssim  10\,{\rm PeV}$. The comparison of the middle and right panels moreover highlights the extreme sensitivity of the parameter $f_{\PBH}$ to $\beta/H_\ast$ due to the exponential dependence in Eq.~\eqref{eq:Pcoll}.
This sensitivity introduces a tuning $\Delta \sim 10^{-4}$ in the model if one aims at achieving $f_{\rm PBH}\sim 1$ and defines the tuning parameter $\Delta$ as
\bea
 \Delta=\textrm{max}_p \Delta_p\simeq \Delta_{\lambda_1}=\left| \frac{d\log f_\PBH}{d\log \lambda_1}\right| \quad \textrm{with} \quad p=\rho,N,\lambda_1 \quad \textrm{and} \quad 
  \Delta_p=\left| \frac{d\log f_\PBH}{d\log p} \right|\,,
\eea 
where the dependence of $f_{\rm PBH}$ on $\lambda_1$ is obtained from Eqs.~\eqref{eq:betaH_Sn_lambda1}, \eqref{eq:Pcoll} and \eqref{eq:fPBH}. The fine-tuning in our model ($\gtrsim 10^{-4}$) turns out to be several orders of magnitude less severe than that required to achieve $f_{\rm PBH}\simeq 1$ in typical single-field inflationary models~\cite{Cole:2023wyx}.
The border line of the gray area in Fig.~\ref{fig:MPBHrho} defines the maximal fraction of PBH, $f_{\rm PBH}^{\rm max}$ (left and middle panels), or the minimal duration of FOPT $(\beta/H_\ast)_{\rm min}$ (right panel), which is experimentally allowed for a monochromatic population of PBH with mass $M_{\rm PBH}$. The black dots along this line correspond to the benchmark points presented in Table \ref{tab:benchmark}, in which we also quote the values of the PBH quantities obtained with the above expressions. Increasing $-\lambda_1$ of a benchmark point would correspond to moving on the vertical dashed line below the black dot of that given benchmark.

\begin{figure}[htb]
  \centering
  \hspace{4mm}\includegraphics[width=5.2cm]
  {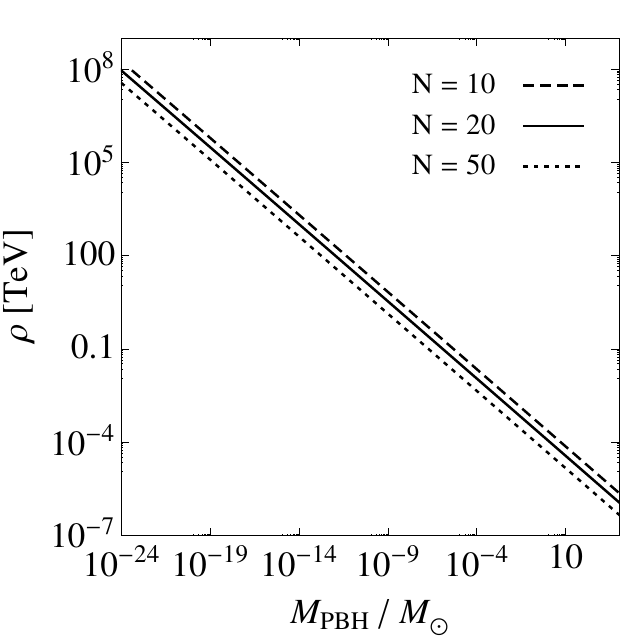}    \hspace{0.2cm}  \hspace{0.2cm}  \includegraphics[width=5.9cm]{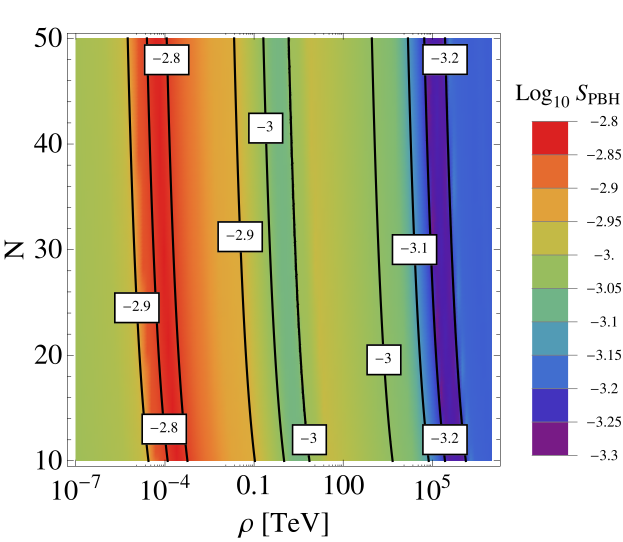}  \\
   \includegraphics[width=9cm]{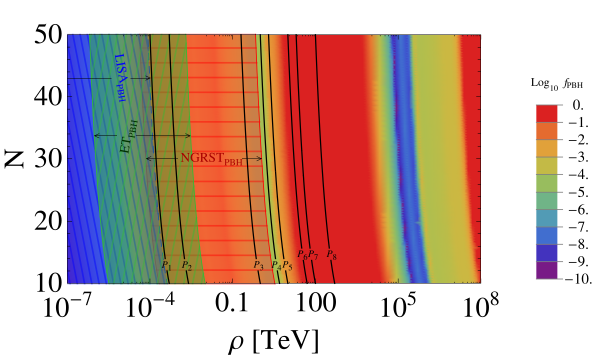}
  \caption{\it \textbf{Upper left panel:} $\rho$ as a function of $M_{\PBH} /
    M_{\odot}$ for  $N=10, 20, 50$ (dashed, solid, dotted line). When varying $M_{\PBH}$, $f_{\rm PBH}$ is adjusted to the maximal value that is experimentally allowed for that value of $M_{\PBH}$ (cf.~border line of the gray area in Fig.~\ref{fig:MPBHrho}). 
    \textbf{Upper right panel:} Contour plot of the BH spin, $S_{\PBH}$, as given by Eq.~(\ref{eq:spin_BH}), in the plane $\{\rho, N\}$.     
    \textbf{Lower panel}: As for the upper right panel but for $f_{\PBH}$. The parameter reach for PBH detection by the LISA, ET, and NGRST experiments is included.
In the upper right and lower panels, we have considered $\beta/H_\ast=(\beta/H_\ast)_{\rm min}$. }
\label{fig:frhobeta}
\end{figure}

Along the border of the gray area in Fig.~\ref{fig:MPBHrho}, the scale $\rho$ varies from 10$^{-7}\,\TeV$ to 10$^{8}\,\TeV$. 
Along this border, $\rho$ depends on $M_{\rm PBH}$ and $N$ as highlighted in the upper left panel of Fig.~\ref{fig:frhobeta}, while the 
dependence on $N$ (for a fixed $f_{\rm PBH}$) is shown in the 
lower panel of this figure where we display the contour lines of $f_\PBH$ in the plane $\{\rho,N\}$ for $\beta/H_\ast = (\beta/H_\ast)_{\rm min}$, i.e.~along the border of the gray area in Fig.~\ref{fig:MPBHrho}.~\footnote{The contour lines highlight the behavior  $f_{\textrm{PBH}}\sim \rho\,N$. This dependence can be understood from Eqs.~\eqref{eq:TRsection} and \eqref{eq:MPBH}, taking into account that $\beta/H$ and $S_n$ vary very little along the gray border in Fig.~\ref{fig:MPBHrho}.} As shown in the figure, parameter regions with $\beta/H_\ast \gtrsim (\beta/H_\ast)_{\rm min}$ fulfill current PBH constraints, but the corresponding PBH signatures are in the reach of LISA, ET, or NGRST if $\rho\lesssim 0.1\,$TeV. 

In the upper right panel of Fig.~\ref{fig:frhobeta} we exhibit contour lines of the PBH initial spin.  
Its value is mild, typically $\sim 10^{-3}$. However, there exist
various mechanisms that can enhance the initial spin of PBHs. For example, PBHs can undergo accretion and gain spin~\cite{DeLuca:2020bjf}, whereas PBH mergers~\cite{Hofmann:2016yih}, or close hyperbolic encounters~\cite{Jaraba:2021ces},
can do the job in convenient astrophysical environments. Moreover, in the presence of beyond the Standard Model localized at the TeV brane, PBHs can spin up 
e.g.~through the emission of light axion or axion-like particles (which could be measured within 10\% by future gamma-ray observatories, provided that $S_\PBH>0.1$)~\cite{Calza:2021czr,Calza:2023rjt}. 
PBHs with spin also exhibit what is known as superradiance~\cite{Yang:2023aak,Tsukada:2018mbp,Berti:2019wnn,Arvanitaki:2010sy}, which could also lead to observable signatures, and modify the initial spin of PBH. Spinning PBHs also may show as primordial features in the SGWB spectrum as investigated, originating from second-order tensors~\cite{Bhaumik:2022zdd} and cosmic strings in the early universe~\cite{Ghoshal:2023sfa}.
Depending on the specific mechanism, the final spin can be enhanced up to two orders of magnitude, from the initial
spin of PBHs created. Moreover, a method for measuring the PBH spin using the peak amplitude of the GW strain has been described in~\cite{Ferguson:2019slp}. For a recent study of PBH spin formation in arbitrary FOPTs, see~\cite{Banerjee:2023qya,Banerjee:2024nkv}.\footnote{In this work, we restrict ourselves to considering only a monochromatic PBH mass
distribution defined by the mass inside the sound horizon, unlike Refs.~\cite{Lewicki:2024ghw,Baldes:2023rqv} which also investigated extended PBH mass distributions. Moreover, for simplicity we do not consider the fluctuation of
the nucleation time of the first bubble, unlike Ref.~\cite{Lewicki:2024ghw} which also takes into account the impact of fluctuation of the nucleation time of the first $j_c$ bubbles. Finally we remark that Refs.~\cite{Flores:2024lng,Lewicki:2023ioy} study PBH formation in purely vacuum-energy dominated patches only, before the beginning of any bubble nucleation. Particularly, Ref.~\cite{Flores:2024lng} singles out this scenario and raises concern that curvature perturbation may get efficiently generated after nucleation of bubbles has started. Presence of such curvature may play a role in modifying the dynamics. In the present work however, we treat late-blooming patches as evolving independently from the background and do not consider the effects of the curvature $K$ that would result from their interaction. Going beyond these assumptions would be interesting but is beyond the scope of the current analysis.}

\section{Gravitational waves}
\label{sec:GW}
\noindent
References~\cite{Megias:2020vek, Megias:2023kiy} detail the FOPT SGWB predictions for the specific high- and low-energy setups of Sec.~\ref{sec:Model}. Here we briefly revisit those computations in view of the new semi-analytic approximations provided in Sec.~\ref{sec:PhaseTransition} and the updates on the SGWB frequency shape recently summarized in Ref.~\cite{Caprini:2024hue}. 

As motivated in Ref.~\cite{Megias:2018sxv}, the FOPTs in our setups are so supercooled ($\alpha \gtrsim 10^7$) that one expects the bubble to expand at the speed of light and the plasma to play a negligible role in the GW production. In this regime where the GW signal is mainly sourced by the bubble collision mechanism, the SGWB 
critical energy density per logarithmic frequency interval is given by~\cite{Caprini:2024hue}
\begin{equation}
h^2\Omega_{\GW}(f) \simeq \frac{16 (f/f_p)^{2.4}}{\left[ 1 + (f/f_p)^{1.2} \right]^4} h^2\bar\Omega_{\GW} \,,
\label{eq:spectrum}
\end{equation}
where $h \equiv 10^{-2} H_0 \,\text{Mpc} /(\text{km/s})  \simeq 0.674$
is the dimensionless parameter of the Hubble constant today, $H_0$, while $h^2 \bar\Omega_{\GW}$ and $f_p$ are respectively the spectrum peak's amplitude and frequency given by
\begin{eqnarray}
  h^2 \bar\Omega_{\GW} &\simeq& 3.8 \times 10^{-6} \left( \frac{H_\ast}{\beta} \frac{\alpha}{1+\alpha} \right)^2 \frac{1}{g_c^{1/3}(T_R)} \,, 
  \label{eq:omegap}\\
  f_p/\textrm{Hz} &\simeq& 8.4 \times 10^{-7}  \frac{\beta}{H_\ast} \frac{T_R \, g_c^{1/6}(T_R)}{0.1 \, \textrm{TeV}} \,.
    \label{eq:fp}
\end{eqnarray}
The SGWB spectrum then exhibits a broken-power-law frequency shape. In particular, at $f \gg f_p$, it behaves as $\Omega_{\GW}(f) \propto f^{-2.4} $ and is much steeper than what Refs.~\cite{Megias:2020vek, Megias:2023kiy} considered (based on the now outdated Ref.~\cite{Caprini:2019egz}), i.e.~$\Omega_{\GW}(f) \propto f^{-1} $. 

\begin{figure}[tb]
  \centering
\includegraphics[width=12cm]{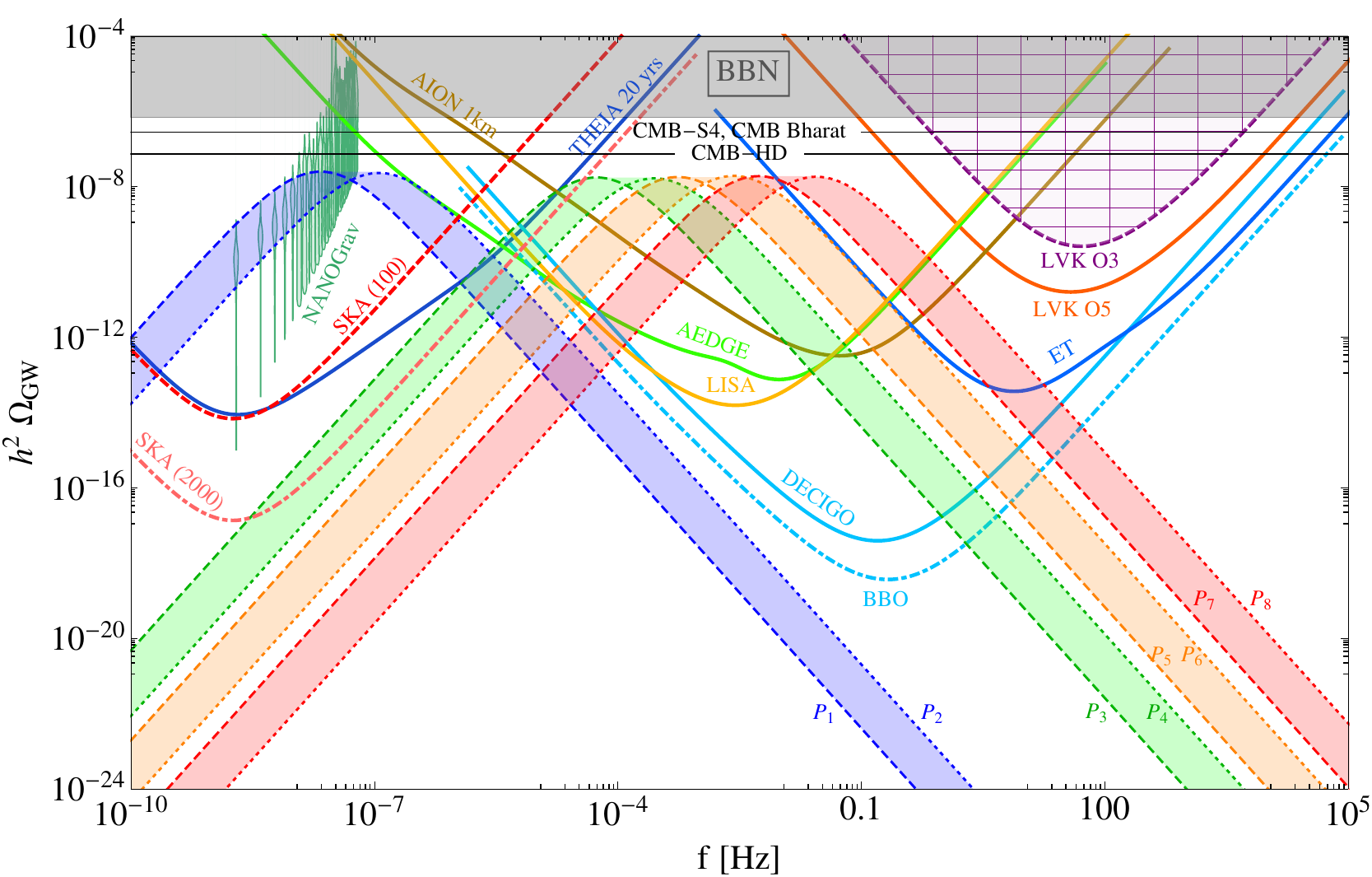}  \\
\caption{\it Current bounds, future sensitivities and SGWBs produced by the FOPTs of the benchmark points in Table~\ref{tab:benchmark}. Dashed and dotted lines on the edge of the strips correspond to benchmarks with $N = 10$ (P${}_{1,3,5,7}$) and $N = 50$ (P${}_{2,4,6,8}$), respectively. The current BBN and LVK bounds are shown in gray and in hashed, while the representative NANOGrav's violin plot illustrates the present PTA measurements. The horizontal lines mark the detection prospects the SGWB signal via its imprint on the relativistic degrees of freedom in the late universe, $N_{\rm eff}$.  The colored ``parabolas" represent the peak-integrated sensitivities of future experiments: a benchmark signal with its peak above the parabola of a given experiment has $\textrm{SNR} > 2$ in that experiment.
}
\label{fig:spectrum}
\end{figure}

\begin{figure}[t]
  \centering
   \hspace{-2cm}\includegraphics[width=7.7cm]{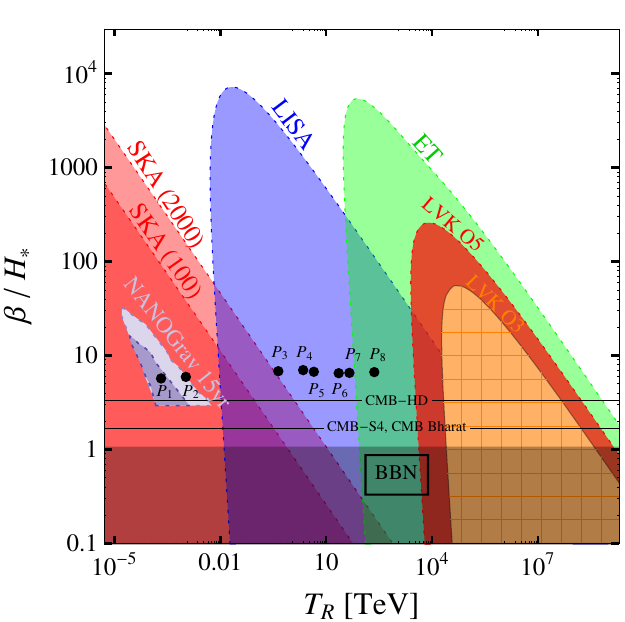}\\ 
 \includegraphics[width=12cm]{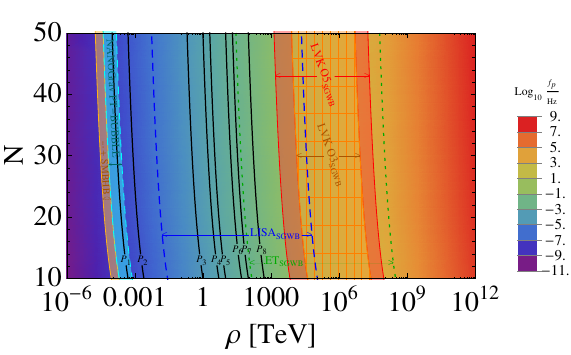}  \\
  \caption{\it 
  \textbf{Upper panel:} The parameter reach of the present and future GW experiment network. In each area the corresponding experiment (see labels) detects the FOPT SGWB with SNR $> 2$. The area of NANOGrav 15yr shows the 95\% C.L.~favored region of Ref.~\cite{NANOGrav:2023hvm}, being the lighter area the region shown in the original reference, while the darker area corresponds to the one shown in the erratum of that reference.  The BBN bound (horizontal gray band) rules out the region $\beta/H_\ast <1.1$. The position of the benchmark points of Table~\ref{tab:benchmark} are shown as black bullet points. In the regions labeled as LISA, ET, LVK O3 and LVK O5, the corresponding experiments detect the FOPT SGWB with SNR $> 2$.
  \textbf{Lower panel:} 
 Contour regions of the peak frequency $f_{p}$ in the plane $\{\rho,N\}$  for $\beta/H_\ast = ( \beta/H_\ast )_{\textrm{min}}$, i.e.~the minimal value allowed by the PBH constraints in Fig.~\ref{fig:MPBHrho}. The parameter reach for SGWB detection at LVK O5, LISA, ET and NANOGrav, along with the excluded region by LVK O3, is derived from the upper panel.}
\label{fig:f_peak}
\end{figure}

We compute $h^2 \bar\Omega_{\GW}$ and $f_p$  arising in our benchmark points P${}_{1}$-P${}_{8}$. Table \ref{tab:benchmark} quotes the results while Figure~\ref{fig:spectrum} displays the corresponding SGWB spectra. Note that, in every pair of benchmark points, P${}_{1}$-P$_{2}$, P${}_{3}$-P$_{4}$, $\dots$, the only different input is $N$, being either 10 or 50 (and a small adjustment in $\beta/H$, c.f.~Table \ref{tab:benchmark}); keeping the same inputs and varying $N$ in the range $10<N<50$ would then move the SGWB signal inside the colored strips in the figure. As expected from Eq.~\eqref{eq:omegap}, $h^2 \bar\Omega_{\GW}$ is essentially constant throughout our benchmark points where $\beta/H$ little varies (and the dependence on $\alpha$ disappears as $\alpha \gg 10$). For the same reason, $f_p$ is mainly controlled by $T_R$ whose dependence on $N$, $\rho$ and $\beta/H$ is discussed in Sec.~\ref{sec:Model}. We see in the lower panel of Fig.~\ref{fig:f_peak} how such a dependence impacts $f_p$ when $\beta/H$ is varied to maximize the PBH abundance that is experimentally allowed at any given PBH mass (c.f.~gray border in the right panel of Fig.~\ref{fig:MPBHrho}). In the same panel, we display the parameter regions that current and forthcoming PTA experiments and interferometers can probe via FOPT SGWB measurements. Such regions are computed as described hereafter.

Experiments can probe a large portion of the parameter space $\{h^2 \bar\Omega_{\GW}, f_p\}$  between now and the far future. The first class of probes leverages the CMB and BBN observations constraining the radiation energy density in the late universe. 
Their constraint can be cast in terms of $\Delta N_{\rm eff}=N_{\rm eff}-N_{\rm eff}^{\rm SM}$, the number of additional degrees of freedom beyond the SM ones that are relativistic at the time of recombination: $\Omega_{N_{\rm eff}} \lesssim 5.6\times10^{-6}\;\Delta N_\text{eff}$, with $N_{\rm eff}^{\rm SM}=3.0440(2)$~\cite{Akita:2020szl,Froustey:2020mcq,Bennett:2020zkv}.
Crucially, as for the energy budget of the universe, the SGWB acts as an extra contribution to the radiation energy density, and the CMB and BBN bounds hold.  For the SGWB spectrum in Eq.~\eqref{eq:spectrum}, this bound implies an upper limit on $h^2 \bar\Omega_{\GW}$:  
\begin{align}
    \int_{f_\text{min}}^{\infty} \frac{\text{d}f}{f}  h^2 \Omega_\text{GW}(f)\simeq 2.22\, h^2\bar\Omega_{\rm GW}  \leq 5.6\times10^{-6}\;\Delta N_\text{eff} \,, \label{eq:darkrad}
\end{align}
where $f_\text{min}\sim 10^{-10}\, \text{Hz}$ for the BBN bound and $f_\text{min}\sim 10^{-18}\, \text{Hz}$ for the CMB bound, but the integral is numerically insensitive to the specific value of $f_\text{min}$ for signals peaking at much higher frequencies. Using the PLANCK 2018 and BBN bound, $N_{\rm eff}=2.99\pm 0.17$~\cite{ParticleDataGroup:2022pth}, we obtain $\Delta N_{\rm eff}\equiv N_{\rm eff}-N_{\rm eff}^{\rm SM}<0.279$  
at 95\% C.L.~and, in turn, $h^2\bar \Omega_\text{GW} \leq  7.2\times 10^{-7}$ at 95\% C.L. 
Moreover, future CMB observations at CMB-S4~\cite{CMB-S4:2020lpa,CMB-S4:2022ght} and CMB-HD~\cite{Sehgal:2019ewc,CMB-HD:2022bsz} are estimated to probe deviations 
equivalent to $\Delta N_{\rm eff}\sim 0.06$ and $\Delta N_{\rm eff}\sim 0.027$ at 95\% C.L., i.e.~$h^2\bar\Omega_{\rm GW}<1.5\times 10^{-7}$ and $h^2\bar\Omega_{\rm GW}<6.8\times 10^{-8}$, respectively.
We display these future reaches and bounds in
Fig.~\ref{fig:spectrum} as a horizontal gray band and horizontal solid lines.

The second class of probes employs GW direct detections. 
Current PTA and LVK measurements fall within this class. The former provides an amplitude range for a SGWB at nanohertz frequencies~\cite{NANOGrav:2023hvm, EPTA:2023fyk, Reardon:2023gzh, Xu:2023wog}. For illustrative purposes, in Fig.~\ref{fig:spectrum} we display the NANOGrav's violin plot as a PTA representative constraint~\cite{NANOGrav:2023hvm} (the violin plots of the other PTA collaborations are similar~\cite{InternationalPulsarTimingArray:2023mzf}). The interpretation of this signal is still under debate and there exist dedicated analyses aiming at scrutinizing the FOPT origin. Here we limit ourselves to qualitative arguments. First, 
we see that the radion FOPTs predicted at the benchmark points P$_1$ and P$_2$ would substantially contribute to the observed PTA data. Nevertheless, to fully fit the violin plot, more power is required at $f\simeq 10^{-7}$\,Hz. This may (likely) arise from a population of individually-unresolved supermassive BH binaries. Instead, within the radion FOPT picture only, this would require to move away from  P$_1$ and P$_2$ by increasing both $h^2 \bar\Omega_{\GW}$, and $f_p$, i.e.~increasing $\rho$ and decreasing $\beta/H_\ast$ (or, equivalently, decreasing $|\lambda_1|$). This however violates the PBH bound, unless the production mechanism of Sec.~\ref{sec:PBH} can be circumvented or made less efficient (see e.g.~Ref.~\cite{Hashino:2025fse}). On the other hand, radion FOPT with $T_R$ slightly below $10^{-4}\,\TeV$ are in some tension ($> 95\%$ C.L.) with the PTA data for producing an excess of signal in the lowest-frequency violin posteriors; FOPTs with $T_R\ll 10^{-4}$\,TeV would be as good as those with  $T_R\gg 10^{-3}$\,TeV as their SGWB would be negligible in the PTA frequency band.

The LVK collaboration provides a 95\% C.L.~upper bound on the SGWB at decihertz frequencies based on the O3 Run~\cite{KAGRA:2021kbb}. Their 95\% C.L.~bound constrains power-law SGWB signals reaching signal-to-noise ratio $\text{SNR} \geq 2$ over one year of data taking, approximately. We recast this bound to a SGWB with the broken-power-law structure in eq.~\eqref{eq:spectrum} 
by constructing the SNR\,$= 2$ ``peak integrated sensitivity" \cite{Schmitz:2020syl}. We thus determine the parameter region in the plane $\{f_p, h^2 \bar \Omega_{\rm GW} \}$ yielding
\begin{align}
     \text{SNR}(f_p, h^2 \bar \Omega_{\rm GW} ) \equiv \sqrt{\tau_i \int_{0}^{\infty} \text{d}f \left(\frac{ h^2 \Omega_\text{GW}(f)}{h^2 \Omega_{\text{exp},i}(f)}\right)^2 } > 2  \,, \label{eq:SNR}
\end{align}
where $\tau_i$ is the length of the data in seconds, and $h^2 \Omega_{\text{exp},i}(f)$ is the experiment's noise curve parametrized as an energy density.\footnote{We remind that $\Omega_{\text{exp}}(f)$ is related to the noise power spectral density $S_h$ by the equality 
$ \Omega_{\text{exp}}(f)  = \left[(2\pi^2 f^3)/ (3 H_0^2)\right]\, S_h(f)$\,, and to the (dimensionless) noise characteristic strain $h_c(f)=\sqrt{f S_h}$ by the equality $\Omega_{\text{exp}}(f)  = \left[(2\pi^2 f^2)/ (3 H_0^2)\right]\, h_c^2(f)$.}
Using $\tau_{\text{LVK O3}} = 11\,$months and $ h^2 \Omega_{\text{exp,LVK O3}}$ provided in Ref.~\cite{KAGRA:2021kbb}, we obtain the ``LVK O3" hatched area in Fig.~\ref{fig:spectrum}. By construction, any SGWB with the frequency shape in Eq.~\eqref{eq:spectrum} is ruled out at 95\% C.L.~if its peak is above the lower border of this area.

With the same procedure, we forecast the FOPT SGWB detection reach of future GW experiments. Specifically, we take the expected noise curves for aLIGO design (from Fig.~1 of Ref.~\cite{LIGOO5}), ET (from Fig.~6 of Ref.~\cite{Hild:2010id}, ET-D curve), AION-km (from Fig.~1 in Ref.~\cite{Badurina:2021rgt}), AEDGE (from Fig.~2 in Ref.~\cite{Badurina:2021rgt}), BBO (from Fig.~3 of Ref.~\cite{Yagi:2011wg}), DECIGO (from Fig.~3 of Ref.~\cite{Yagi:2011wg}), LISA (from Fig.~1 in Ref.~\cite{LISA:2017pwj}), $\mu$-ARES (from Fig.~1 in Ref.~\cite{Sesana:2019vho}), THEIA (from Eq.~(2.3) in Ref.~\cite{Garcia-Bellido:2021zgu}) and SKA (from Ref.~\cite{Moore:2014lga, MooreWebpage} with 100 and 200 millisecond pulsars as input), 
assume for concreteness $\tau_{\rm aLIGO des.}=\tau_{\rm ET}= \tau_{\rm AION\, km} =  \tau_{\rm AEDGE} =  \tau_{\rm LISA} =  \tau_{\rm \mu-ARES} = 4\,$yr and $\tau_{\rm THEIA}=\tau_{\rm SKA}=20\,$yr, and finally compute the regions of the plane $\{f_p, h^2 \bar \Omega_{\rm GW} \}$ satisfying the condition in Eq.~\eqref{eq:SNR}. The lower borders of these regions are presented in Fig.~\ref{fig:spectrum}, and a FOPT with frequency shape as in Eq.~\eqref{eq:spectrum} with the peak above a given border would have SNR\,$>\,2$ in the corresponding detector. The synergy and complementarity of the considered GW experiments is manifest  in Fig.~\ref{fig:spectrum}.

It is worth stressing that a SGWB source with a given SNR in a detector implies that its signal is present in the data with some statistical significance. Nevertheless, the SNR criterion does not guarantee detection. Foregrounds due to unresolved individual sources may hinder the detection of even SGWBs with high SNR. Moreover, for signal-dominated experiments where the noise curve must be determined together with the signals, the SNR evaluation does not include the uncertainties on the noise reconstruction. To be precise on the detection capabilities, one should run the data analysis pipeline for the SGWB reconstruction data analysis pipeline dedicated to every experiment (see e.g.~Refs.~\cite{Badger:2022nwo, Gowling:2022pzb, Hindmarsh:2024ttn, Caprini:2024hue} for proof-of-principle examples). Given the complexity of such a task, we limit ourselves to the SNR estimate and let the scientific collaborations reach more quantitative conclusions.

Finally, in the upper panel of Fig.~\ref{fig:f_peak} we remap the current bounds and forthcoming probes as functions of $T_R$ and $\beta/H$. To show the GW experiment synergy expected in the late 40s, we focus on the detection capabilities (based on the simplistic proxy SNR\,$>\,2$) of ET, aLIGO design, LISA and SKA with $\tau_\textrm{aLIGO des.} = \tau_\textrm{ET} = 7\,$yr, $\tau_\textrm{LISA} = 4\,$yr and $\tau_\textrm{SKA} = 20\,$yr. It results that the future network of GW experiments has a huge indirect discovery power on PBHs, as their sensitivities reach radion FOPTs that lead to PBH abundances of only $f_{\rm PBH} \gtrsim 5 \times  10^{-6}$. In particular, the region where PTA data likely require a SGWB is displayed in the light gray and dark gray areas corresponding respectively to Ref.~\cite{NANOGrav:2023hvm} and erratum.\footnote{We display the NANOGrav result for concreteness. Other PTA experiments obtain essentially the same result~\cite{InternationalPulsarTimingArray:2023mzf}.} We see that the benchmark point P$_1$ is inside this region, so the values of the (low-energy) 5D setup parameters favored by the PTA measurement should be around those of P$_1$, in agreement with Ref.~\cite{Megias:2023kiy}. In this remapping, the current BBN and Planck constraint displayed in Fig.~\ref{fig:spectrum} rules out the region $\beta/H_\ast < 1.1$ at 95\% C.L., while the future CMB-S4 and CMB-HD observations will further test the region $1.1 < \beta/H_\ast < 2.3$ and $1.1 < \beta/H_\ast < 3.4$ at 95\% C.L..

\section{Collider and astrophysical data}
\label{sec:collider}
\noindent
The universal feature of 5D models is the presence of KK gravitons and scalar (radion) fluctuations propagating along the fifth dimension. As explained above, we are assuming the SM propagating on the TeV (or SM) brane and so the coupling of gravitons with mass $m_h^n = c_n \rho$ and the radion (localized toward the IR brane) depends on the distance between the SM brane and the IR brane. In such a scenario, the phenomenology is driven by the first graviton KK mode with mass $m_h = c_1 \rho  \equiv \hat m_h \rho$ ($\hat m_h \simeq 3.83$) and the lightest radion state with mass $m_\chi \equiv \hat m_\chi \rho$ ($\hat m_\chi \simeq 0.01 - 0.1$), which have wave functions $h_{\mu\nu}(x)$ and $\chi(x)$, respectively.\footnote{We are not considering the massless graviton zero mode which is very weakly coupled to the SM as its interaction is suppressed by~$1/M_P$.} These fields are coupled to the SM with Lagrangian~\cite{Csaki:2000zn}
\be
\mathcal L=-\kappa_h h_{\mu\nu}(x) T^{\mu\nu}_{\rm SM}(x) -  \kappa_\chi \chi(x)  \eta_{\mu\nu} T^{\mu\nu}_{\rm SM}(x) \,,
\ee
where $T^{\mu\nu}_{\rm SM}(x)$ is the stress-energy tensor. The couplings $\kappa_h$ and $\kappa_\chi$ in terms of our parameters $\rho$ and $N$ are then given by
\begin{align}
\kappa_h &= \frac{2\sqrt{2}\pi}{N}\frac{1}{\rho}\,\left[\Theta\left(\frac{\rho-\rho_T}{\textrm{TeV}}\right)+\left( \frac{\rho}{\rho_T}\right)^2 \frac{J_2(c_1 \rho/\rho_T)}{J_2(c_1)} \Theta\left(\frac{\rho_T-\rho}{\textrm{TeV}}\right)\right]  \,,\\
\kappa_\chi &= \frac{2\pi}{\sqrt{3}N}\frac{1}{\rho}\,\left[\Theta\left(\frac{\rho-\rho_T}{\textrm{TeV}}\right)+\left( \frac{\rho}{\rho_T}\right)^2\Theta\left(\frac{\rho_T-\rho}{\textrm{TeV}}\right)\right]  \,, 
\end{align}
where $J_n(x)$ is a Bessel function of the first kind, and $\Theta(x)$ is the Heaviside step function equal to 1 (0) for $x\geq 0$ ($x<0$) which allows us to automatically swap from the high-energy to the low-energy setup when we cross the threshold $\rho =1\,$TeV. (As explained in Sec.~\ref{sec:Model}, we choose such a value of $\rho$ for concreteness, but other values of the order of 1 TeV would work as well.)

\subsection{Bounds from the KK graviton sector}
\label{subsec:bounds_KK}
\begin{figure}[tb]
  \centering
    \includegraphics[width=5.0cm]{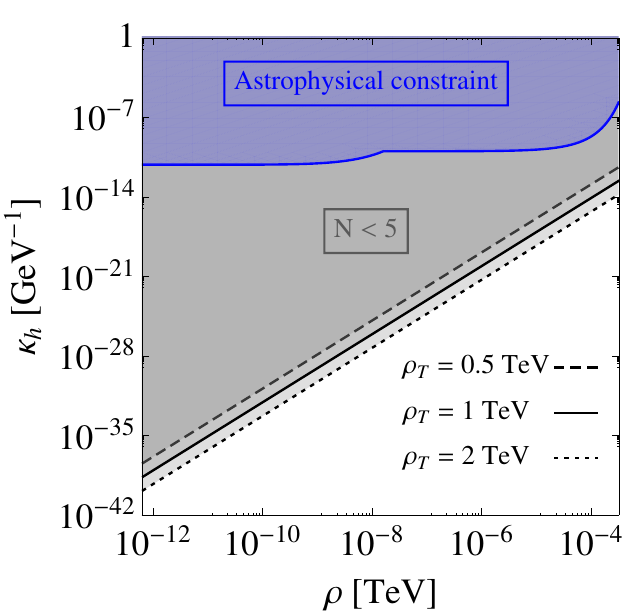} \hspace{0.2cm}  
    \includegraphics[width=4.9cm]{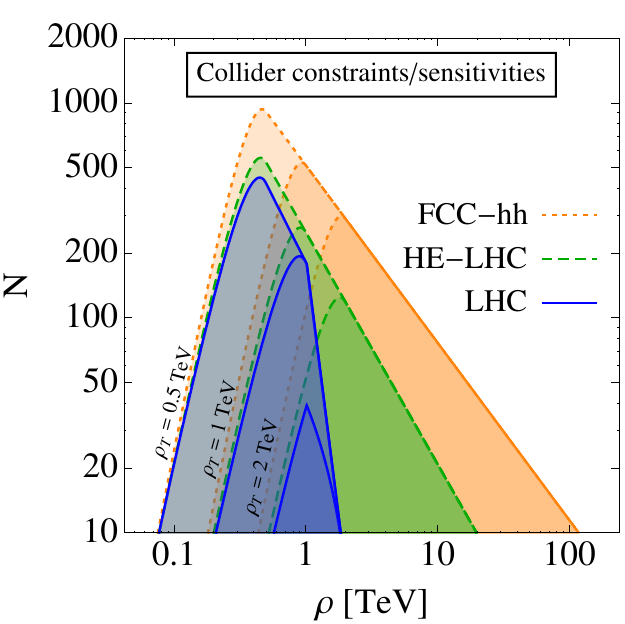}  \hspace{0.2cm}
    \includegraphics[width=4.5cm]{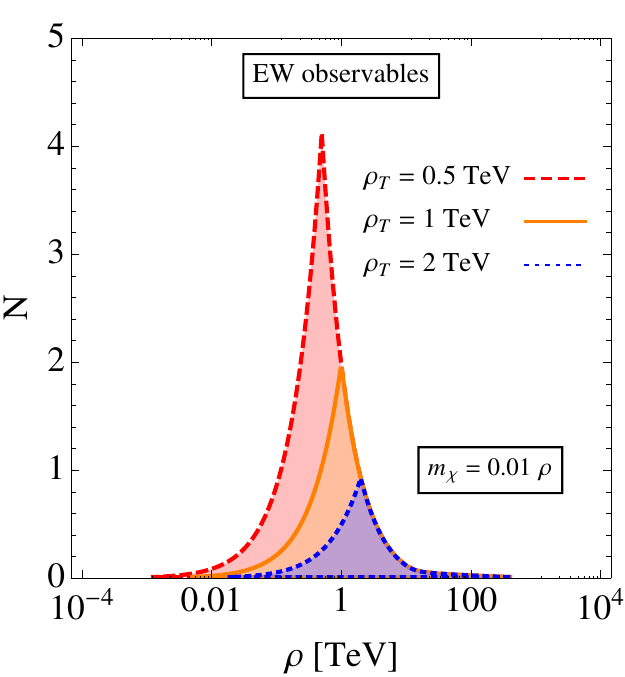} \\
  \caption{\it  \textbf{Left panel:} The parameter regions ruled out by astrophysical observations (blue area)~\cite{Cembranos:2017vgi, Cembranos:2021vdv} and breaking of perturbativity bound $N\gtrsim 5$ (gray area)~\cite{Agashe:2007zd,Chen:2014oha} for several values of $\rho_T \sim 1$\,TeV. 
  \textbf{Middle panel:} The LHC bound and future-collider sensitivity reaches on the plane $\{N,\rho\}$ for $\rho_T=0.5,\,1,\,2$ TeV. The LHC, HE-LHC and FCC-hh rule out the blue, green and orange areas, respectively. The LHC bound is based on  $\sqrt{s}=13$ TeV and 139\,fb$^{-1}$ data~\cite{ATLAS:2021uiz,ATLAS:2024fdw}. The HE-LHC and FCC projections assume 100 \,ab$^{-1}$ data at $\sqrt{s}=27$\,TeV and $\sqrt{s}=100$\,TeV, respectively~\cite{Helsens:2019bfw,Harris:2022kls}. \textbf{Right panel:} Constraint in the plane $\{\rho,N\}$ coming from the EW observables $S$ and $T$ for $\rho_T = 0.5 ,\, 1,\,2$\,TeV and  $m_\chi=0.01\,\rho$. The shaded areas are excluded at the 95\% CL.
  }
\label{fig:astro-collider}
\end{figure}

\noindent
The existence of the KK graviton resonances influences how the strength of the gravitational force scales with distance and how fast the star's cores lose energy due to the additional emission channel~\cite{Cembranos:2017vgi, Cembranos:2021vdv}. Astrophysical data thus constrain the mass and couplings of the KK graviton states (see Fig.~3 in Ref.~\cite{Cembranos:2021vdv}). In the left panel of Fig.~\ref{fig:astro-collider}  (blue area), we present the corresponding bound on $\kappa_h$ once the graviton mass is expressed in terms of $\rho_T$ and $N$.  In the same panel, we display the upper bound (straight diagonal lines) due to the perturbativity constraint $N \gtrsim 5$~\cite{Agashe:2007zd,Chen:2014oha}  for several choices of $\rho_T\sim \mathcal O (1\,\textrm{TeV})$. 
Since the astrophysical constraint rules out a region inside the area where the perturbativy limit is violated, the astrophysical experiments do not impose any stringent constraint on the model.

In the middle panel of Fig.~\ref{fig:astro-collider}, we present the parameter reach achieved, or expected to be achieved, at current or future colliders~\cite{Cembranos:2021vdv,ATLAS:2022hwc,ATLAS:2024fdw}. For $\rho_T=0.5, 1,2\,$TeV, the null searches for the graviton KK modes at the LHC, for $\sqrt{s}=13$\,TeV and 139\,fb$^{-1}$ luminosity, rule out the corresponding blue regions. At 
 $\rho \sim \rho_T$, LHC imposes a strong lower bound on $N$. Still for high enough values of $N$ experimental results can be evaded as the coupling behaves as $\kappa_h\propto 1/N$. Analogous searches at HE-LHC and FCC-hh will expand the probed area, with a major improvement in sensitivity toward much higher values of $\rho$. In particular, for the coupling $k/M_P=0.1$ and a luminosity of 100\,ab$^{-1}$, the HE-LHC at $\sqrt{s}=27$\,TeV will probe the region $m_h \lesssim 10$\, TeV at 95\% C.L., while the LHC-hh at $\sqrt{s}=100$\,TeV will probe the region  $m_h \lesssim 100$\, TeV~\cite{Helsens:2019bfw,Harris:2022kls}.

\subsection{Bounds from the light radion sector}
\noindent
The (light) radion interacts with the Higgs doublet $H=(0,v+h)^T/\sqrt{2}$ due to the interaction term~\cite{Folgado:2019sgz}
\begin{equation}
\mathcal L \supset  -\frac{1}{2}\lambda_{\chi hh} \chi(x) h^2(x)  \,,   \qquad \lambda_{\chi hh} =\frac{8\pi}{\sqrt{3}N}m_H^2 \left[\frac{1}{\rho}\,
\Theta\left(\frac{\rho-\rho_T}{\textrm{TeV}}\right)+\frac{\rho}{\rho_T^2}\,
\Theta\left(\frac{\rho_T-\rho}{\textrm{TeV}}\right)  \right]  \,,
\end{equation}
where $m_H$ is the SM Higgs mass and we use the Heaviside function to swap from the low-energy to high-energy setup when $\rho$ crosses 1\,TeV.
Due to this interaction, the radion appears in the electroweak precision observables. We study its impact by computing its contribution to the parameters $S$ and $T$~\cite{Csaki:2000zn,Gunion:2003px} and comparing it with the $\chi^2$ ellipse parametrization, leading to values compatible with EW observables within 95\% C.L.. The outcome is presented in the right panel of Fig.~\ref{fig:astro-collider} for different value of $\rho_T$ and $m_\chi=0.01\,\rho$ (the outcome is similar for $m_\chi = 0.1 \rho$).
Due to the small coupling of the radion with the SM, the parameter space in the plane $\{\rho,N\}$ is very mildly excluded by EW observables. The exclusion mainly happens for values  $\rho \sim 1\, \TeV$, but it occurs at very small values of $N$ $(N \lesssim 5)$. This is beyond the range of reliability of the model (cf.~left panel of Fig.~\ref{fig:astro-collider}), as we are neglecting gravitational quantum corrections in our treatment of the 5D model.

\begin{figure}[t]
  \centering
     \includegraphics[width=7cm]{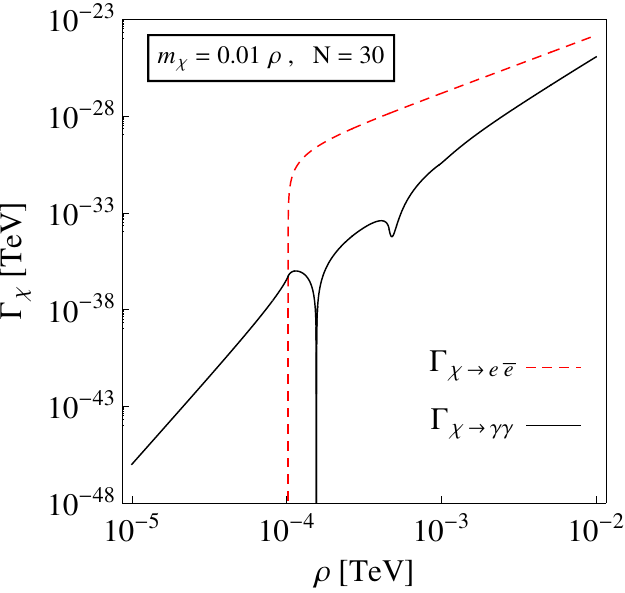} \hspace{0.5cm}  \includegraphics[width=6.5cm]{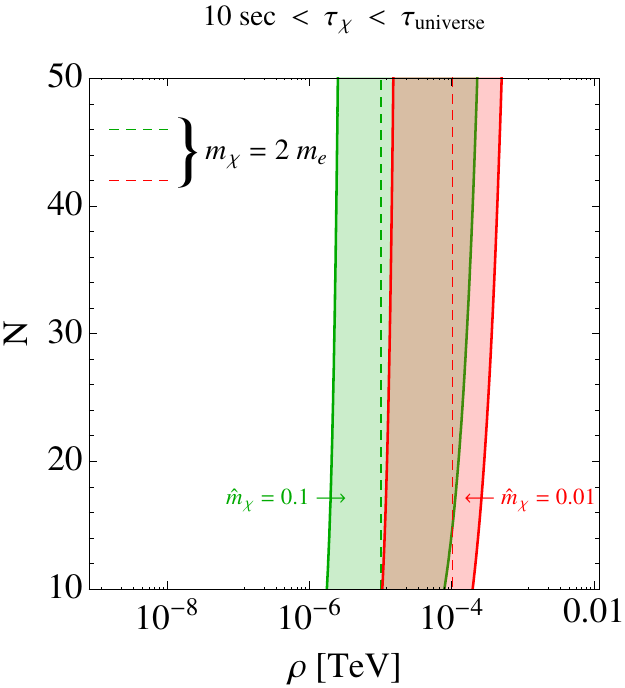}  \\
     \caption{\it \textbf{Left panel:}  Decay width of the radion as a function of $\rho$. We display the two contributions: $\Gamma_{\chi \to \gamma\gamma}$ (solid black) and $\Gamma_{\chi \to e \bar e}$ (dashed red). We have considered $\hat m_\chi \equiv m_\chi / \rho = 0.01$ and $N = 30$. \textbf{Right panel:} Excluded band by BBN due to the radion decay, $10 \, \textrm{sec} < \tau_\chi < \tau_{\rm universe}$. 
   The allowed (white) region corresponds to $\tau_\chi<10$ sec ($\tau_\chi>\tau_{\rm universe}$) on the right-side (left-side) of the forbidden band.   
     We display the results for $\hat m_\chi = 0.1$ (green region) and $\hat m_\chi = 0.01$ (red region). We display also as dashed vertical lines the value $\rho = 2 m_e / \hat m_\chi$, corresponding to $m_\chi = 2 m_e$.}
\label{fig:taurbound}
\end{figure}

Finally, we will study the constraint related to the radion decay into SM fields, and triggered by BBN data, for our low-energy setup which assumes $\rho<\rho_T$. The radion can decay into photons, gluons and all the fermion channels. The radion width for the decay in the channel $\chi \to \gamma \gamma$ reads as
\begin{equation}
\Gamma_{\chi \to \gamma\gamma} = \frac{\alpha_{\QED}^2 b_{\QED}^2}{192 \pi} \frac{m_\chi^3 \rho^2}{N^2 \rho_T^4} \,,
\end{equation}
while the decay in the channel $\chi \to ee$ is given by
\begin{equation}
\Gamma_{\chi \to e \bar e} = \frac{\pi}{6} \frac{m_\chi m_e^2 \rho^2}{N^2 \rho_T^4}  \left( 1 - \frac{4 m_e^2}{m_\chi^2}\right)^{3/2} \,.
\end{equation}
These are the only relevant channels if we focus in the region $m_\chi < 2 m_\mu$, and  $\Gamma_{\chi \to e \bar e}$ is the dominant one  for $2 m_e \lesssim m_\chi < 2 m_\mu$.  Then the total width is $\Gamma_\chi \simeq \Gamma_{\chi \to \gamma \gamma} + \Gamma_{\chi \to e\bar e}$, and the radion lifetime $\tau_\chi = 1 / \Gamma_\chi$ turns out to be~\cite{Koutroulis:2024wjl}
\begin{equation}
\tau_\chi  \simeq \left\{ 
\begin{array}{lc}
8.4 \times 10^{-44} \, \textrm{sec} \times N^2 \left( \frac{\rho}{\textrm{TeV}} \right)^{-2} \left( \frac{m_\chi}{\TeV}\right)^{-7}    &\quad \textrm{for} \quad  m_\chi < 2 m_e \\
 4.8 \times 10^{-15}  \, \textrm{sec} \times N^2 \left( \frac{\rho}{\TeV} \right)^{-2} \left( \frac{m_\chi}{\TeV} \right)^{-1} \left(1 - \frac{4 m_e^2}{m_\chi^2} \right)^{-3/2}  &\quad\textrm{for}\quad  2 m_e < m_\chi
\end{array}
\right.    \,.    \label{eq:tau_r_ana}
\end{equation}
In this expression we have set $\rho_T = 1 \, \TeV$, and used $b_{\QED} \simeq \frac{7}{90} \frac{m_\chi^2}{m_e^2}$ for $m_\chi < 2 m_e$.  We display  the widths  $\Gamma_{\chi \to \gamma\gamma}$ and $\Gamma_{\chi \to e \bar e}$ in the left panel of Fig.~\ref{fig:taurbound}.

A constraint follows from the fact that the radion decays would photodissociate the BBN elements if $\tau_\chi > \tau_{\rm BBN} \approx 10\, \textrm{s}$~\cite{Abu-Ajamieh:2017khi}, and then such a long lifetime would be forbidden. This is so except for $\tau_\chi > \tau_{\textrm{universe}}$, as in this case the radion can be considered stable. The lifetime that is actually excluded is then  
\begin{equation}
\tau_{\textrm{universe} }>\tau_\chi  >10 \, \textrm{s} \,,
\end{equation}
which translates into a forbidden window in the space $\{N,\rho\}$ by using Eq.~(\ref{eq:tau_r_ana}) with some given values of $m_\chi$. This excluded region is shown in the left panel of Fig.~\ref{fig:taurbound} and rules out  $2  \, \textrm{MeV} \lesssim \rho \lesssim 0.1 \, \textrm{GeV}$  and $ 10 \, \textrm{MeV} \lesssim \rho \lesssim 0.3  \, \textrm{GeV}$ for $m_\chi = 0.1 \rho$ and $m_\chi = 0.01 \rho$, respectively, which are natural choices for the radion mass~\cite{Megias:2020vek}. The bound then varies with different choices of the model parameters that affect the radion mass. 

Note that the parameter regions for which the radion lifetime fulfills the condition $\tau_\chi > \tau_{\rm universe}$ or barely overcomes the bound $\tau_\chi < \tau_{\rm BBN}$, remain questionable. In Sec.~\ref{sec:PhaseTransition}, we compute the thermodynamic properties of the phase transition 
for which
nucleation and percolation occur close to each other, and we can assume that reheating is practically instantaneous after percolation. A very long radion lifetime jeopardizes this approximation in the corresponding (small) region of the parameter space. Reaching firm conclusions in this regime would require more advanced techniques than those employed in this paper.

\section{Summary results and conclusions}
\label{sec:conclusions}
\noindent
Warped extra dimensional models are among the most promising candidates for solving, or at least alleviating, the gauge hierarchy problem of the Standard Model (SM). Additionally, they naturally exhibit a strong first-order phase transition (FOPT), a key ingredient for explaining the baryon asymmetry of the universe (BAU). However, the dark matter (DM) puzzle does not allow for an easy solution: the warped embedding does not offer a stable particle meeting the DM constraints unless additional, ad-hoc ingredients are invoked~\cite{Panico:2008bx,Carena:2009yt,Koutroulis:2024wjl}. In this paper, we have demonstrated that this is not the case, as the unavoidable FOPT arising in warped models can also produce the required DM observables in the form of primordial black holes (PBHs). The satisfactory parameter window is small but appealing, as it predicts several BSM signatures that fall within the reach of multiple future experiments.

To quantitatively demonstrate the viability of the PBH DM option in a 5D embedding, we have considered two warped extradimensional setups. The first setup is a Randall-Sundrum (RS) model with a ultraviolet (UV) and an infrared (IR) brane, with the SM localized on the latter. The second setup is an RS generalization with an additional intermediate brane (between the UV and IR branes) hosting the SM fields. In both setups, the energy scale of the IR brane is set by a radion field that acquires a vacuum expectation value (VEV) $\rho$ and breaks conformal invariance. Alleviating the hierarchy problem requires $\rho \gtrsim 1\,$TeV in the first setup, and $\rho \lesssim 1\,$TeV in the second setup. Consequently, in our study, we have amply covered the theoretically-reasonable range of $\rho$ by analyzing the interval $10^{-7}\,\TeV < \rho < 10^{8}\,\TeV$, assuming either the first (for $\rho \ge 1\,\TeV$) or the second setup (for $\rho < 1\,\TeV$).

Since the SM fields are localized on a brane, their interactions are SM-like. The BSM signatures and bounds arise from the Kaluza-Klein (KK) gravitons and radions propagating in the bulk. Regarding the graviton, we have computed its coupling to the SM stress-energy tensor and identified the parameter space ruled out by astrophysical observations, perturbativity requirements, and LHC data. Additionally, we have determined the parameter region within the reach of the High-Luminosity LHC (HL-LHC) and Future Circular Collider (FCC). Concerning the radion, we have computed its lifetime and decay channels, and identified the parameter space where these observables are compatible with Big Bang Nucleosynthesis (BBN). We have also studied the (supercooled) FOPT triggered by the radion, and the PBHs and stochastic gravitational wave background (SGWB) that such a phase transition generates. To carry out a systematic analysis with limited computational resources, we have derived some semi-analytic expressions improving the thick-wall approximation (which turns out to be quite rough in estimating the nucleation temperature) and avoided the fully numerical computations for the bounce.

\begin{figure}[tb]
  \centering
  \includegraphics[width=12cm]{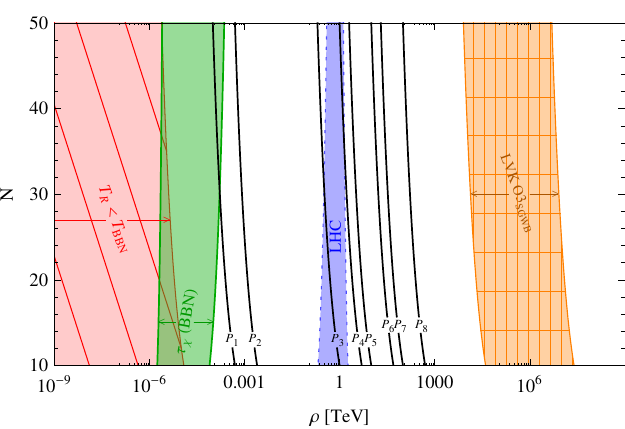}   \\
  \caption{\it The regions in the plane $\{\rho, N\}$ forbidden by present SGWB searches, collider analyses on direct detection of KK gravitons, and astrophysical and cosmological observations. The bounds are based on the assumption that, at every value of $\rho$, the PBH abundance is maximized to its experimental upper limit. For other assumptions, see the sections where the bounds are derived.}
\label{fig:final}
\end{figure}

\begin{figure}[t]
  \centering
      \includegraphics[width=12cm]{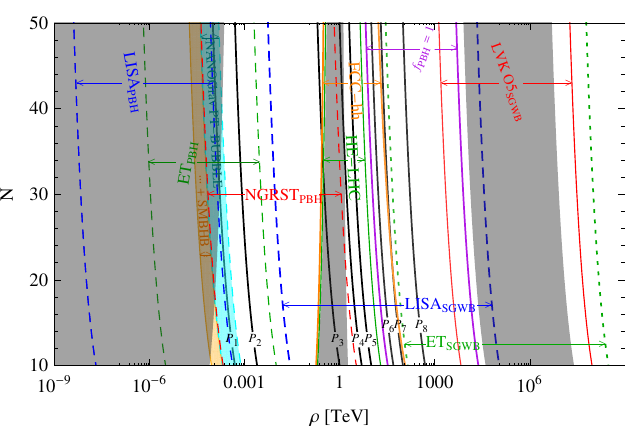}   \\ 
  \caption{\it Parameter regions probed by GW interferometers via the PBH SGWB (LISA$_\textrm{PBH}$, ET$_\textrm{PBH}$) and FOPT SGWB (LISA$_\textrm{SGWB}$, ET$_\textrm{SGWB}$, LVK\,O5$_\textrm{SGWB}$), by microlensing observations (NGRST$_\textrm{PBH}$), and collider searches (HL-LHC, FCC-hh). Dark areas are forbidden according to Fig.~\ref{fig:final}. PBH abundance matches the observed DM abundance in the region $f_{\textrm{PBH}}=1$.}
\label{fig:final2}
\end{figure}

In Figs.~\ref{fig:final} and \ref{fig:final2} we highlight the complementarity and synergy among PBH observations, SGWB probes, and collider tests. The two figures provide an overall view of the results detailed in the previous sections, as well as our benchmarks points P$_1$-P$_8$ (see Table~\ref{tab:benchmark}). We recall that for every value of $\rho$ and $N$, the inverse FOPT duration $\beta/H_\ast$ is fixed such that the quantity $f_{\rm PBH}$, defined as the ratio of PBH abundance to the total dark matter abundance, reaches its experimental upper limit (see Fig.~\ref{fig:MPBHrho}); much larger values of $\beta/H_\ast$ would relax the displayed bounds. As for the radion mass parameter $\hat m_\chi$ appearing in our collider and astrophysics computations, the choice $\hat m_\chi = 0.1$ is assumed, but other reasonable values would only mildly modify the results. The FOPT SGWB sensitivity regions are based on a signal-to-noise ratio (SNR) threshold SNR $> 2$, which neglects the subtleties related to the existence of foregrounds, loud transient signals, and instrumental noise uncertainties. 
 Last but not least, our estimates of the PBH abundance triggered by FOPTs rely on recent literature and are consequently evolving fast. Repeating the analysis of the present paper in more depth will surely be worthwhile when the PBH and FOPT SGWB predictions settle down.

Bearing in mind the above caveats, Fig.~\ref{fig:final} singles out the parameter space that is currently ruled out. The parameter region labelled as ``$\tau_\chi \; (\textrm{BBN})$" and ``$T_R< T_{\rm BBN}$", corresponding to $\rho\lesssim 10^{-4}$, lead to either a radion lifetime or a FOPT entropy injection problematic for BBN and Planck observations. The region with   $\rho\sim 1$\,TeV conflicts with the outcome of the existing LHC searches. The LVK O3 run rules out the region $10^4 \lesssim \rho/{\rm TeV} \lesssim 10^7$.

On the other hand, Fig.~\ref{fig:final2} shows the overall excluded region together with the parameter reach of future colliders, gravitational-wave (GW) observatories and microlensing experiments, aiming at the detection of the KK resonances, FOPT SGWBs or PBH signatures.
We conclude that the region with $f_{\rm PBH}=1$, which corresponds to $\rho \sim 1 - 1000\,$TeV (with the lower value in tension with LHC data), will be probed with LVK O5, and subsequently tested by LISA and ET via the FOPT SGWB observables, and later on by FCC-hh searches for graviton resonances. Given the multitude of literature's BSM model proposals predicting a FOPT SGWB, the synergy between colliders and GW detectors will be of paramount importance for scrutinizing whether our warped setups are the right ones responsible for the phenomenology that will be observed. Finally, SGWB searches at current and future ground-based interferometers will probe the radion FOPT that maximizes the PBH abundance at $\rho \sim 10^4 - 10^7\,$TeV. At these or higher scales, however, the warped setups partly loose theoretical motivation as they do not solve the full  hierarchy problem, and the PBH abundance is orders of magnitude below the DM one (see Fig.~\ref{fig:MPBHrho}).

As Fig.~\ref{fig:final2} shows, multiple experiments will also test the region $\rho < 1\,$TeV for which the additional intermediate brane is required. In this region, the PBH constraints forbid $f_{\rm PBH}=1$. However, the maximal abundance experimentally allowed at $\rho \sim 10^{-3} - 1\,$TeV and not yet ruled out by the current LHC bound on the graviton resonances (due to the suppressed coupling of KK gravitons to the TeV brane) will be within the reach of microlensing experiments such as the Nancy Grace Roman Space Telescope (NGRST). Moreover, the narrow parameter region around our benchmark point P$_1$ provides a good explanation of the SGWB signal hint that current pulsar timing array (PTA) experiments are finding. This region will be further probed by ET and LISA by measuring the SGWB due to the unresolved PBH population that the radion FOPT predicts.

This landscape of both synergetic and complementary probes presents a unique opportunity for model selection, i.e.~the identification of a (hopefully small) set of BSM setups compatible with the observed signatures. A generic prediction of the warped setup we have investigated is the occurrence of a highly supercooled radion-driven FOPT characterized by a large strength parameter, $\alpha \gg 1$.

For such large values of $\alpha$, the SGWB is expected to be dominated by the bubble-collision contribution, with a frequency spectrum distinct from that generated by sound waves or turbulence, which tend to dominate when $\alpha \lesssim 1$. At the same time, for $\alpha\gg 1$ and $\beta/H_* \lesssim 10 $, which maximizes the PBH abundance, the SGWB signal is so loud that it can be precisely reconstructed provided its peak is well within an interferometer sensitivity band (see e.g.~Fig.~4 in Ref.~\cite{Caprini:2024hue}). Taken together, these effects suggest that if, for instance, one of our benchmarks P$_4$-P$_8$ is realized in nature, future GW interferometers will be able to favor setups with a large supercooling while ruling out the plethora of BSM scenarios that are either incapable of producing such a strong FOPT, or that produce it in the sound-wave or turbulance regime.

Moreover, supercooling generically implies low values of $\beta/H_*$, which enables a sizeble PBH production during the FOPT. In fact, as we can see from Fig.~\ref{fig:final}, a positive SGWB detection by an interferometer should be accompanied by PBH production, which could be detected by a different experiment.  For instance, parameter scenarios close to benchmarks P$_2$ and P$_3$ would be detected by both FOPT SGWB searches at LISA and microlensing PBH imprints at  NGRST. Another example is the parameter region around the benchmark P$_1$. This region would explain the current PTA SGWB hint both in the presence and absence of a SGWB contribution due to supermassive black hole binaries. In addition, it would lead to a PBH SGWB signal detectable at ET, and PBH microlensing effects testable at NGRST.

Of course the main discriminator of the warped model compared to other BSM exhibiting strong FOPTs is the discovery of the graviton KK modes by collider experiments. For instance, the LISA discovery
of the FOPT SGWB predicted around benchmarks P$_4$ and P$_5$  (with peak frequency $f_p\sim (3-6)\times 10^{-4}$\,Hz and amplitude $h^2 \bar\Omega_{\rm GW}\sim 10^{-8}$) would be accompanied by detection of graviton KK modes with masses $m_{\rm KK}$ in the multi-TeV range at the future HE-LHC. Similarly, the LISA discovery of a FOPT SGWB related to the parameter region between the benchmarks P$_6$ and P$_7$ (with peak frequency $f_p\sim (0.6-3)\times 10^{-3}$\,Hz and amplitude $h^2\bar\Omega_{\rm GW}\sim 10^{-8}$) should be followed by discovery of graviton KK modes with masses $m_{\rm KK}\sim 10^2$ TeV at the future FCC-hh collider.  However, for the case of the PTA SGWB detection, the graviton KK modes with masses around or below the GeV scale are too weakly coupled to ordinary matter to be tested at present or future colliders.

\begin{acknowledgments}
  \noindent

EM would like to thank the IFAE, Barcelona, Spain, for hospitality during the
final stages of this work.  The work of EM is supported by the project
PID2020-114767GB-I00 funded by MCIN/AEI/10.13039/501100011033, and by
the Junta de Andaluc\'{\i}a under Grant FQM-225. GN is partly
supported by the grant Project.~No.~302640 funded by the Research
Council of Norway. The work of MQ is supported by the grant
PID2023-146686NB-C31 funded by MICIU/AEI/10.13039/501100011033/ and by
FEDER, EU. IFAE is partially funded by the CERCA program of the
Generalitat de Catalunya.

\end{acknowledgments}

\appendix

\section{Thick-wall approximation}
\label{sec:AppA}
\noindent 
In this appendix, we derive the temperatures characterizing the phase transition. As we are interested in the large supercooling regime, we restrict ourselves to the $O(4)$ symmetric solution to the bounce equation in the thick-wall approximation. Within this approximation, the action of the bounce solution can be estimated as~\cite{Megias:2021rgh}
\begin{equation}
S_4(T) \simeq \frac{9 N^2}{4 \mathcal V(T/\rho)} \qquad \textrm{with} \qquad \mathcal V(T/\rho) = \pi^4 (T_c^4 - T^4) / \rho^4 \,.
\end{equation}

Bubbles start to form below the critical temperature $T_c$ at which the free energies of the RS and BH phases are equal. At the temperature $T'<T_c$, the number of bubbles per horizon volume reads as
\be
N_{\rm B}(T')=\int_{T'}^{T_c}\frac{dT}{T}\frac{\Gamma(T)}{H^4(T)}\,,
\label{eq:NB}
\ee
with $\Gamma$ being the $O(4)$-symmetry tunneling rate and the Hubble parameter $H$ given by~\footnote{The vacuum dominated condition is guaranteed as long as $15 N^2 / (4 g_\star) \gg 1$, and it is fulfilled for the typical value $N \gtrsim 10$. For $N = 10$, this ratio becomes $15 N^2 / (4 g_\star) \simeq 3.5$.} 
\begin{equation}
H^2 = \frac{1}{3 M_P^2} \left( \rho_R + \rho_V \right) = \frac{1}{3 M_P^2} \left(  \frac{\pi^2}{30}g_\star T^4 + E_0 \right) \,.
\label{eq:Hubble}
\end{equation}

The nucleation temperature $T_n$ is defined as the temperature yielding $N_{\rm B}(T_n)=1$. The supercooling regime is characterized by $T_n \ll T_c$, implying the Hubble parameter to be dominated by the vacuum energy and to have negligible $T$ dependence. Consequently, with the change of integration variable $T\to S_4(T)$ in eq.~\eqref{eq:NB}, we can rewrite the condition $N_{\rm B}(T_n)=1$ as
\be
\frac{\pi^2}{N^4}  \left(\frac{M_P}{\rho} \right)^4  S_0 \left[ e^{-S_n} - S_0 e^{-S_0} \textrm{Ei}(S_0 - S_n)\right]=1 \,,
\label{eq:Sn_1}
\ee
where 
\begin{eqnarray}
S_0\equiv S_4(0) \,,\\
S_n\equiv S_4(T_n) \,.
\label{eq:Sn_def}
\end{eqnarray}
and $\textrm{Ei}(x) \equiv \int_{-\infty}^x \frac{e^t}{t} dt$ (with $x < 0$) is the exponential integral function. Thanks to eq.~\eqref{eq:betaH}, $S_0$ and $S_n$ can be related as
\be
S_0\simeq S_n-\frac{1}{4}\, \frac{\beta}{H_\ast}  \,,
\ee
which allows us to simplify eq.~\eqref{eq:Sn_1} as
\be
 \frac{\pi^2}{N^4}  \left(\frac{M_P}{\rho} \right)^4  S_n e^{-S_n} \left[ 1+ S_n f(\beta/H_\ast)\right]=1,
  \label{eq:Snnumerical}
\ee
where
\be
f(\beta/H_\ast)=-e^{\frac{\beta}{4H_\ast}} \textrm{Ei}\left(-\frac{\beta}{4H_\ast}\right)  \,.
\ee
Since $f(x)$ varies between $f(1)\simeq 1.34$ and $f(10)\simeq 0.3$, the term $S_n f(\beta/H_\ast)$ in Eq.~\eqref{eq:Snnumerical} is the leading one for $\beta/H_\ast\lesssim 10$ and $S_n\gtrsim \mathcal O(10^2)$. Equation~\eqref{eq:Snnumerical} consequently reduces to
\be
S_n^2 \, e^{-S_n}=\frac{N^4}{\pi^2}(\rho/M_P)^4\frac{1}{f(\beta/H_\ast)} \,,
\ee
from where we obtain
\be
S_n=-2\mathcal W_{-1}\left[ -\frac{N^2}{2\pi}\left(\frac{\rho}{M_P} \right)^2 f^{-1/2}(\beta/H_\ast) \right]  \simeq 4 \log\left[ \frac{2\sqrt{\pi}}{N} \left( \frac{H_\ast}{\beta} \right)^{1/4}   \frac{M_P}{\rho} \right] \,.  \label{eq:Sn} 
\ee
Here $\mathcal W_{k}$ is the $k$-th branch of the Lambert function.  The approximation follows from $f(z) \stackrel[z \gg 1]{}{\simeq} \frac{4}{z}$ and $\mathcal W_{-1}(-z) \stackrel[z \ll 1]{}{\simeq} \log z$.

\begin{figure}[t]
  \centering
  \includegraphics[width=7.5cm]{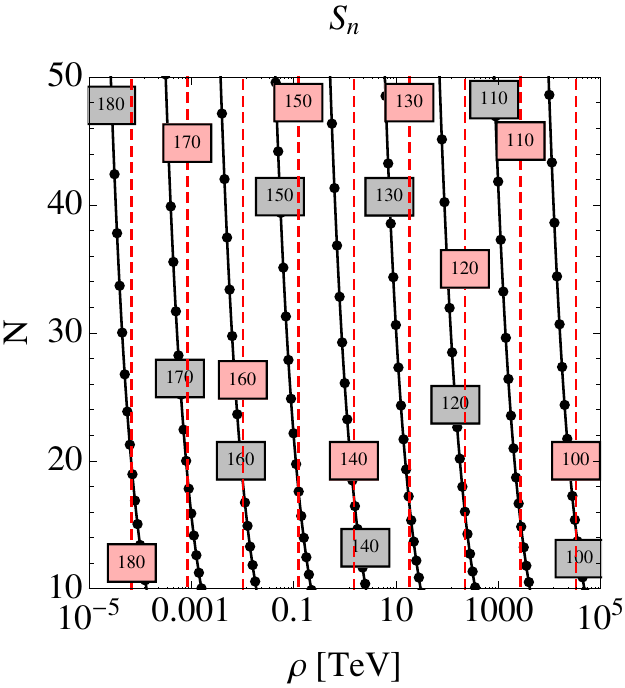} 
  \caption{\it Contour plot of $S_n$ for $\beta/H_\ast=6$. Dots are the numerical solution to Eq.~(\ref{eq:Snnumerical}), solid (black) lines are explicit approximate solution in the first equality of Eq.~(\ref{eq:Sn}), and dashed (red) lines are the function $S_n=4\log(M_P/\rho)$. }
\label{fig:Sn}
\end{figure}

Fig.~\ref{fig:Sn} shows in the plane $\{\rho,N\}$ the mismatch among the different approximations to evaluate $S_n$. For the black dots, $S_n$ is computed numerically from Eq.~(\ref{eq:Snnumerical}); for the solid black curve $S_n$ is approximated as in Eq.~(\ref{eq:Sn}); and red lines correspond to the approximation $S_n=4\log(M_P/\rho)$ rather customary in the literature.
As we can see, the typical values of $S_n$ are $100 \lesssim S_n \lesssim 180$, and the approximation in Eq.~(\ref{eq:Sn}) is small enough not to be visible. The disagreement from the naive estimation $S_n = 4 \log(M_P/\rho)$ is instead visible.

Within the thick-wall approximation, we can achieve some analytical estimates for the critical, nucleation and reheat temperatures in our model, as functions of $\rho$, $N$ and $\beta/H_\ast \lesssim 10$. 
From eqs.~\eqref{eq:betaH},  \eqref{eq:Sn_def} and \eqref{eq:Sn} we extract 
\begin{align}
T_c &=\sqrt{\frac{3N}{S_n}} \left(4S_n+\frac{\beta}{H_\ast} \right)^{1/4}\frac{\rho}{2\pi} \,,  \nonumber \\
T_n &=  \sqrt{\frac{3N}{S_n}} \left(\frac{\beta}{H_\ast} \right)^{1/4} \frac{\rho}{2\pi} \,.   
\end{align}
Moreover, by assuming an instantaneous reheating where all the energy of the potential is converted into radiation~\footnote{A naive condition for instantaneous reheating, that $\tau_\chi\lesssim t_p$, where $t_p$ is the age of the universe at percolation, leads to $\rho\gtrsim O(1)$ GeV, for $m_\chi=0.1 \rho$. Smaller values of $\rho$ should be analyzed in more detail, which is beyond the scope of the present paper.}, and neglecting the nucleation temperature due to the supercooled regime, we compute the reheating temperature
\begin{align}
T_R&\simeq \left(\frac{15N^2}{4g_\ast} \right)^{1/4}T_c  \,. 
\label{eq:TR}
\end{align}

Using the above results, we compute the percolation temperature $T_p$. The bubble nucleation rate is
\begin{equation}
\Gamma(T) \simeq \frac{1}{R_0^4} \left( \frac{S_4}{2\pi}\right)^2 e^{-S_4} \,,
\end{equation}
where $R_0$ is the size of the nucleating bubble, which in the thick-wall approximation is given by
\begin{equation}
R_0 = \frac{\sqrt{6} \rho}{\pi^2 \sqrt{T_c^4 - T^4}}\simeq \frac{\sqrt{6} \rho}{\pi^2T_c^2} \,,\quad \textrm{for} \ T \ll T_c   \,.
\end{equation}
The probability that a given spatial point is still in the false vacuum is given by $P(T) = e^{-I(T)}$, where
\begin{equation}
I(T) = \frac{4\pi}{3} \int_{T}^{T_c} \frac{dT^\prime}{T^{\prime \, 4}} \frac{\Gamma(T^\prime)}{H(T^\prime)} \left[ \int_{T}^{T^\prime} \frac{d\tilde T}{H(\tilde T)} \right]^3  \,. \label{eq:IT}
\end{equation}
One can estimate $T_p$ as the temperature satisfying the condition $I(T_p) = 0.34$ which corresponds to $P \simeq 0.7$~\cite{Ellis:2018mja,Lewicki:2021xku}.
By plugging Eq.~\eqref{eq:Hubble} into Eq.~(\ref{eq:IT}) in the vacuum dominated regime,   the expression for $I(T)$ can be written as
\begin{equation}
I(x) \simeq \frac{972 N}{\pi^{13}} \frac{M_P^4}{R_0^4 \rho^8} \frac{1}{x_c^6}  \int_{x}^{x_c} dx^\prime e^{- \frac{9 N^2}{4\pi^4 ( x_c^4 - x^{\prime\, 4} )} } \frac{(x^\prime - x)^3}{x^{\prime\, 4} ( x_c^4 - x^{\prime\, 4})^2  \sqrt{N^2 x_c^4 + \frac{4}{15} g_\star x^{\prime \, 4} } }  \,, \label{eq:ISTT}
\end{equation}
where $x \equiv T/\rho$ and $x_c \equiv T_c/\rho$.
This expression behaves at low temperature as
\begin{equation}
  I(x) \simeq - \frac{972}{\pi^{13}} \frac{M_P^4}{R_0^4 \rho^8} \frac{e^{- \frac{9 N^2}{4 \pi^4 x_c^4}} }{x_c^{16}} \log x \,,  \qquad (x \to 0) \,,
\end{equation}
so that it is logarithmically divergent. This behavior is also true without the vacuum domination assumption. The condition $I(x_p)\gtrsim 0.34$ is therefore guaranteed at some temperature $T_p$.

The above necessary condition seems to not be sufficient for vacuum dominated phase transitions as the false vacuum is inflating, and even if the probability $P(t)$ decreases with time and reaches the required value, it has to decrease faster than the increase in the volume of the space. An additional requirement for the whole universe ending up in the broken phase is then that the physical volume of the false vacuum $\mathcal V_{\rm false}\propto a^3(t) P(t)$ decreases around the percolation temperature. This translates into the additional condition
\be
T\frac{dI(T)}{dT}\equiv -3K(x)<-3   \,,
\label{eq:Kbound}
\ee
i.e.
\begin{equation}
K(x)\equiv  \frac{972 N}{\pi^{13}} \frac{M_P^4}{R_0^4 \rho^8} \frac{x}{x_c^6}  \int_{x}^{x_c} dx^\prime e^{- \frac{9 N^2}{4\pi^4 ( x_c^4 - x^{\prime\, 4} )} } \frac{(x^\prime - x)^2}{x^{\prime\, 4} ( x_c^4 - x^{\prime\, 4})^2  \sqrt{N^2 x_c^4 + \frac{4}{15} g_\star x^{\prime \, 4} } } > 1 \,, \label{eq:KT}
\end{equation}
where, for $T\ll T_c$, we can use the approximation 
 \be
 \frac{M_P^4}{R_0^4\rho^8}\simeq \frac{\pi^8 x_c^8}{36}\ (M_P/\rho)^4\,.
 \ee
The function $K(x)$ goes to a constant value 
in the limit $x\to 0$:
 \be
 K(x)\simeq I(x) \frac{-1}{3 \log x} \to \frac{9}{\pi^{5}} \frac{M_P^4}{\rho^4} \frac{1}{x_c^8} e^{-\frac{9 N^2}{4 \pi^4 x_c^4 }}  \simeq -\frac{16\pi}{9\,\textrm{Ei}\left(-\frac{\beta}{4H_\ast}\right)}\gtrsim 1 \qquad \textrm{for} \qquad \frac{\beta}{H_\ast}\gtrsim 0.01 \,.
 \ee
 Remarkably, one can see that the condition $K(x)>1$ in eq.~\eqref{eq:Kbound} is not guaranteed for $I(x)=0.34$. The completion of the phase transition then requires that the two following conditions are both satisfied:
 \be
 I(x_p)\geq 0.34 \quad  \&\quad K(x_p)\geq 1 \,.   \label{eq:IKxp}
 \ee

\begin{figure}[t]
  \centering
    \includegraphics[width=5.5cm]{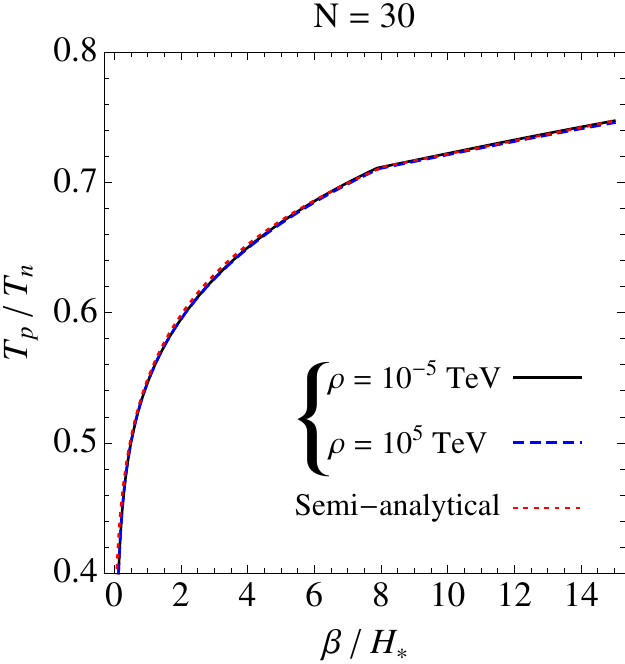} \hspace{1cm}   \includegraphics[width=5.8cm]{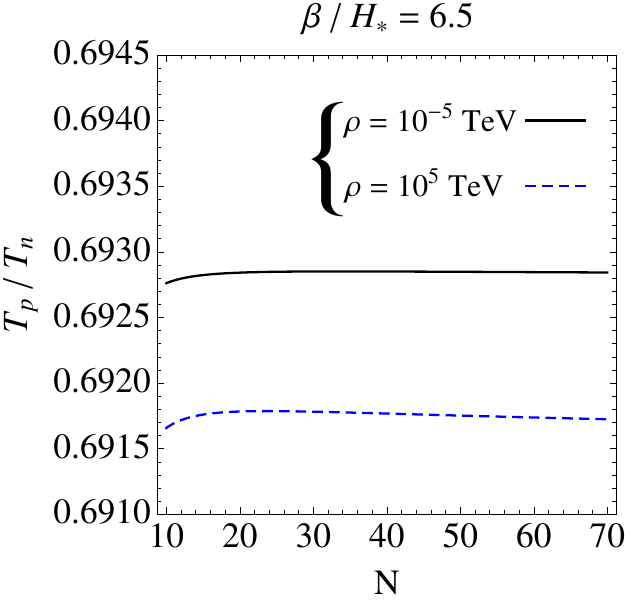}    \\ 
  \caption{\it \textbf{Left panel:} Plot of $T_p/T_n$ as a function of $\beta/H_\ast$ for values of $10^{-5}\leq \rho \leq 10^5$ TeV and $N=30$. \textbf{Right panel:} The same as a function of $N$ for $\beta/H_\ast=6.5$. The solid (black) and dashed (blue) lines are computed by using Eq.~(\ref{eq:IKxp}), while the (red) dotted line in the left panel corresponds to Eq.~(\ref{eq:fitTp}).}
\label{fig:TpTn}
\end{figure}

Figure~\ref{fig:TpTn} shows the dependence of the percolation temperature (relative to the nucleation temperature) on $\rho$, $N$, and $\beta/H_\ast$. We find that the ratio $T_p/T_n$ depends primarily on $\beta/H_\ast$, while its dependence on $\rho$ and $N$ is tiny. Moreover, $I(x_p) = 0.34$ guarantees $K(x_p) > 1$ for $\beta/H_\ast \gtrsim 8$, while $K(x_p) = 1$ leads to $I(x_p) > 0.34$ for $\beta/H_\ast \lesssim 8$, independently of the values of $\rho$ and $N$. 
The dependence on $\beta/H_\ast$ can be fitted with the semi-analytical expression
\be
\frac{T_p}{T_n}=\left\{ 
\begin{array}{lc}
0.548 \left(\frac{\beta}{H_\ast}\right)^{1/8} &\quad \textrm{for}\quad\frac{\beta}{H_\ast}\lesssim 8 \\
0.671+5.1 \times 10^{-3}\left(\frac{\beta}{H_\ast}\right)&\quad\textrm{for}\quad  \frac{\beta}{H_\ast}\gtrsim 8
\end{array}
\right.   \,. \label{eq:fitTp}
\ee
The relationship $T_p < T_n$ always occurs, as expected.

\section{Semi-analytical approximations}
\label{sec:AppB}

In this appendix, we compare the results derived from the thick-wall approximation in Appendix \ref{sec:AppA} with the exact results obtained by solving numerically the bounce equation corresponding to the $O(4)$ symmetry. This comparison allows us to improve the approximation and achieve some semi-analytical, thick-wall inspired, formul$\ae$ for the different phase transition temperatures. These formul$\ae$ are those used in the numerical analysis of this paper, largely speeding up our numerical computations. 

\begin{figure}[t]
  \centering

    \includegraphics[width=6cm]{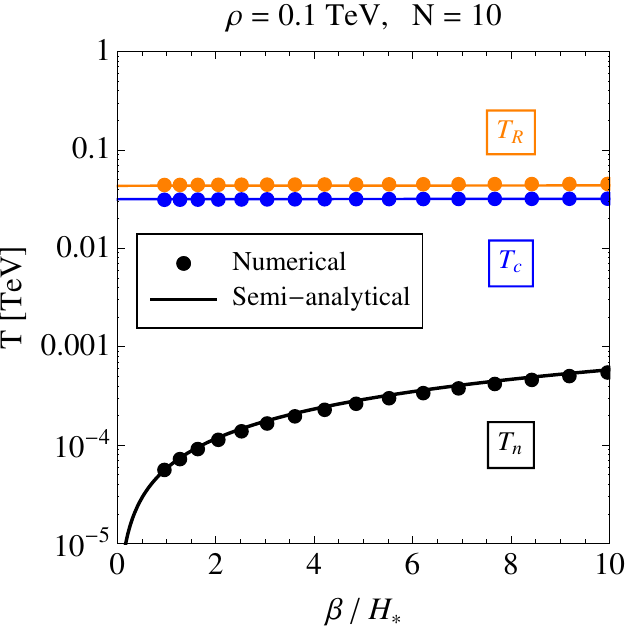}  \hspace{1cm}
    \includegraphics[width=6cm]{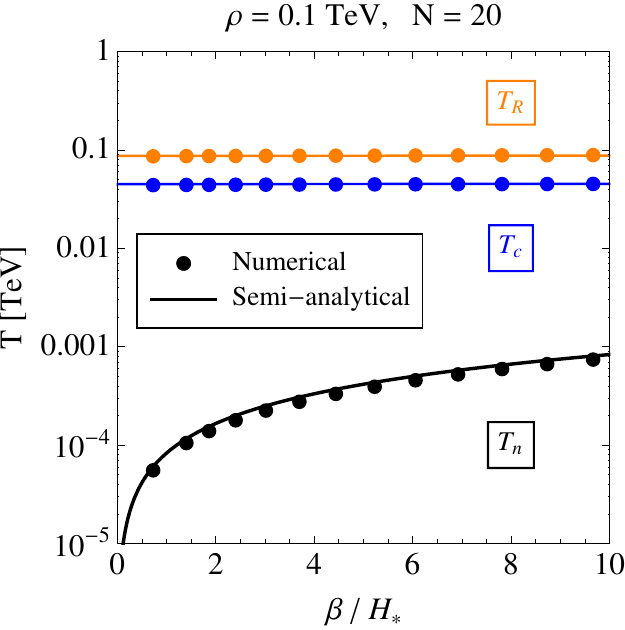}  \\
\includegraphics[width=6cm]{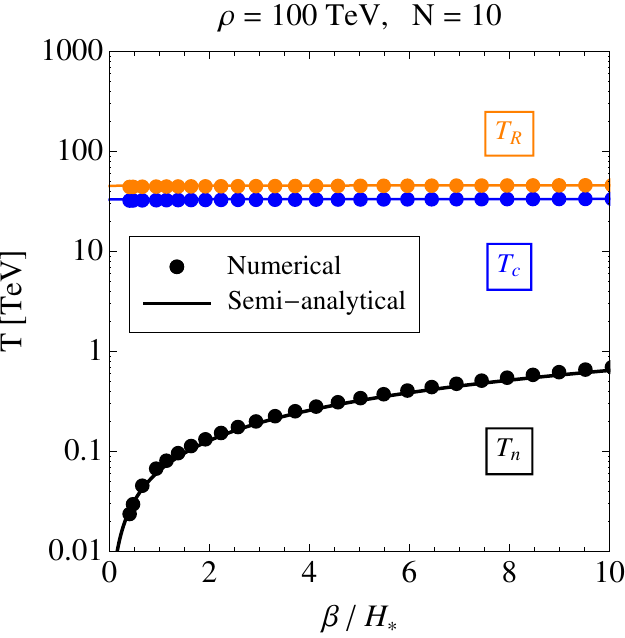} \hspace{1cm}
\includegraphics[width=6cm]{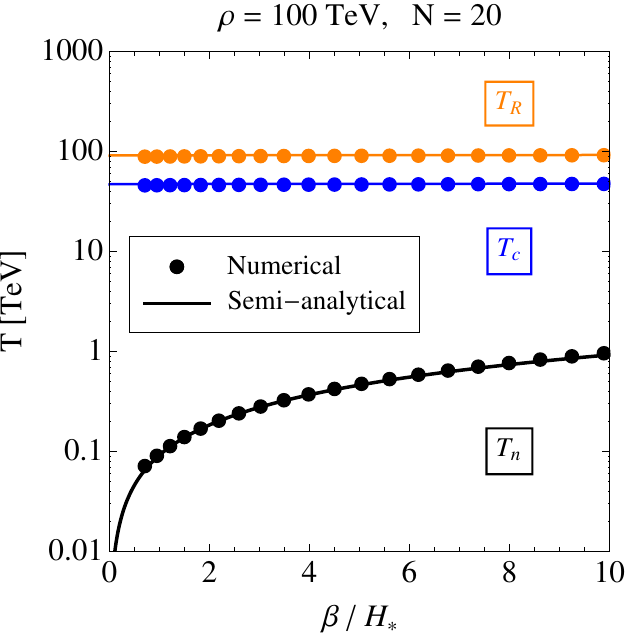}   \\
  \caption{\it Comparison between the temperatures obtained from the exact numerical computation of the bounce equation (dots) and the semi-analytical formul$\ae$ (solid lines) for the temperatures $T_n$, $T_c$ and $T_R$, given by Eqs.~(\ref{eq:Tc_fit})-(\ref{eq:TR_fit}).}
\label{fig:TnumTfit}
\end{figure}

The dots in Fig.~\ref{fig:TnumTfit} show
the values of $T_c$, $T_n$ and $T_R$ obtained from the numerical bounce solution as functions of $\beta/H_\ast$ for $\rho = 0.1$ TeV (upper panels) and $\rho = 100$ TeV (lower panels), and for $N=10$ (left panels) and $N=20$ (right panels). 
As the thick-wall approximation results turn out to be rather off, the figure omits them. It instead includes the semi-analytic approximations (solid lines) based on fitting the temperature values as
\begin{align}
T_c &=a_c\sqrt{\frac{3N}{S_n}}   \left(4S_n+\frac{\beta}{H_\ast} \right)^{1/4}\frac{\rho}{2\pi} \,, \label{eq:Tc_fit}\\
T_n &= a_n \sqrt{\frac{3N}{S_n}}  \left(\frac{\beta}{H_\ast} \right)^{b_n}  \frac{\rho}{2\pi} \,,  \label{eq:Tn_fit}  \\
T_R &\simeq a_R \frac{3^{3/4}5^{1/4}}{2\sqrt{2}\pi} \frac{ \left(4S_n+\frac{\beta}{H_\ast} \right)^{1/4}}{g_\ast^{1/4}\sqrt{S_n}} \, N \rho  \,,  \label{eq:TR_fit}
\end{align}
with
\be
a_c \simeq 0.9 \,, \quad a_n\simeq 8.25 \times 10^{-3} \,,\quad b_n\simeq 1 \,, \quad a_R \simeq 0.9 \,.    \label{eq:ac_aR}
\ee
As Fig.~\ref{fig:TnumTfit} shows, the semi-analytic approximations in Eqs.~(\ref{eq:Tc_fit})-(\ref{eq:TR_fit}) accurately reproduce the fully numerical results in the big-bubble regime $\beta/H_\ast \lesssim 10$ we are interested in. 

We have checked that such semi-analytical approximations also work well for other values of $\rho$ and $N$ within the interval values considered in the present analysis. Equations.~(\ref{eq:Tc_fit})-(\ref{eq:ac_aR}) and (\ref{eq:fitTp}) are then the explicit expressions for the critical, nucleation, and reheating temperatures implemented in the main analysis of this work.

\bibliographystyle{JHEP}
\bibliography{refs}

\end{document}